\documentclass[aps,twocolumn,amsmath,amssymb,superscriptaddress,notitlepage,longbibliography]{revtex4-1}
\usepackage{CJKutf8}
\usepackage[colorlinks=true,citecolor=blue,linkcolor=blue,hypertexnames=false]{hyperref}
\usepackage{amsmath}
\usepackage{dsfont, ulem}
\usepackage{dcolumn}
\usepackage{hyperref}
\usepackage{tikz}
\usepackage{textcomp}
\usepackage{float}
\usepackage{mwe}
\usepackage{graphicx}
\usepackage{ulem}
\usepackage{multirow}

\def \k{{\bm k}}
\def \q{{\bm q}}

\newcommand*{\ket}[1]{|#1\rangle}

\usepackage{siunitx}
\usepackage{bm} 
\usepackage{braket}
\usepackage[version=4]{mhchem}

\newcommand{\old}[1]{}

\DeclareMathOperator{\Tr}{Tr}

\newcommand{\n}[1]{\left| #1 \right|}
\newcommand{\st}[1]{\left\{#1\right\}}
\renewcommand{\v}[1]{\boldsymbol{#1}}

\newcommand{\PRLsec}[1]{\textit{#1} --- }

\newcommand{\eg}{\varepsilon_{\mathrm{Gr}}}
\newcommand{\qIKS}{\v{q}_{\mathrm{IKS}}}

\newcommand{\qIKSy}{q_{\mathrm{IKS}}^y}
\newcommand{\CIKS}{C_{\mathrm{IKS}}}
\newcommand{\xiIKS}{\xi_{\mathrm{IKS}}}
\newcommand{\qCSS}{\v{q}_{\mathrm{CSS}}}

\begin{document}
\begin{CJK*}{UTF8}{bsmi}
\title{Kekul\'e spiral order in magic-angle graphene: a density matrix renormalization group study}
\author{Tianle Wang}
\affiliation{Department of Physics, University of California, Berkeley, CA 94720, USA}
\affiliation{Materials Science Division, Lawrence Berkeley National Laboratory, Berkeley, California 94720, USA}
\author{Daniel E. Parker}
\affiliation{Department of Physics, Harvard University, Cambridge, MA. 02139, USA}
\author{Tomohiro Soejima (副島智大)}
\affiliation{Department of Physics, University of California, Berkeley, CA 94720, USA}
\author{Johannes Hauschild}
\affiliation{Department of Physics, Technische Universit\"at M\"unchen, 85748 Garching, Germany}
\author{Sajant Anand}
\affiliation{Department of Physics, University of California, Berkeley, CA 94720, USA}
\author{Nick Bultinck}
\affiliation{Rudolf Peierls Centre for Theoretical Physics, University of Oxford, Oxford OX1 3PU, United Kingdom}
\affiliation{Department of Physics, Ghent University, 9000 Ghent, Belgium}
\author{Michael P. Zaletel}
\affiliation{Department of Physics, University of California, Berkeley, CA 94720, USA}
\affiliation{Materials Science Division, Lawrence Berkeley National Laboratory, Berkeley, California 94720, USA}

\begin{abstract}

When the two layers of a twisted  moir\'e system are subject to different degrees of strain, the effect is amplified by the inverse twist angle, e.g., by a factor of 50 in magic angle twisted bilayer graphene (TBG). Samples of TBG typically have heterostrains of $0.1-0.7\%$,  increasing the bandwidth of the ``flat'' bands by as much as tenfold, placing TBG in an intermediate coupling regime. Here we study the phase diagram of TBG in the presence of heterostrain with unbiased, large-scale density matrix renormalization group calculations (bond dimension $\chi=24576$), including all spin and valley degrees of freedom. Working at filling $\nu = -3$, we find a strain of $0.05\%$ drives a transition from a quantized anomalous Hall insulator into an incommensurate-Kekul\'e spiral (IKS) phase. This peculiar order, proposed and studied at mean-field level in Ref. \cite{Kwan2021}, breaks both valley conservation and translation symmetry $\hat{T}$, but preserves a modified translation symmetry $\hat{T}'$ with moir\'e-incommensurate phase modulation. Even higher strains drive the system to a fully symmetric metal.

\end{abstract}
\maketitle
\end{CJK*}

\begin{figure}
    \centering
    \includegraphics[width=0.5\textwidth]{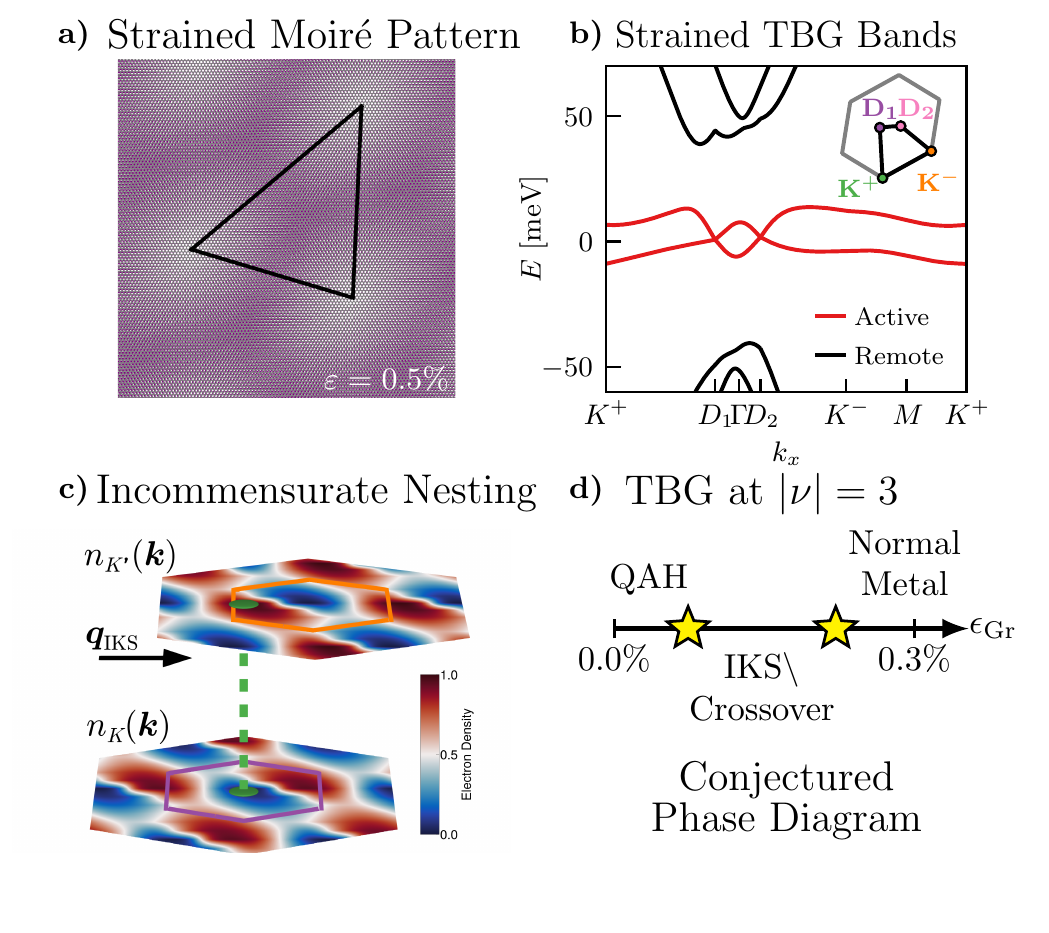}
    \caption{a) Moir\'e pattern from two graphene lattices with $1^\circ$ relative twist and $\eg = 0.5\%$ uniaxial heterostrain. Strain is amplified at the superlattice scale, significantly distorting the moire unit cell. b) Non-interacting bandstructure of TBG with $\eg=0.15\%$ heterostrain. Heterostrain shifts the Dirac nodes $D_{1,2}$ close to the $\Gamma$ point (inset).
    c) Valley-resolved Brillouin zone electron occupations within DMRG in the IKS phase (parameters match Fig.\,\ref{fig:IKS}). The occupations are approximately nested under a relative shift $\v{q}_{\mathrm{IKS}}$; note $\Gamma$ is depleted in both valleys. d) \textit{Conjectured} phase diagram of Eq.~\eqref{eq:interacting_TBG}, TBG with heterostrain, at $|\nu|=3$.
    }
    \label{fig:overview}
\end{figure}

 Strong-coupling theories of magic-angle twisted bilayer graphene (TBG) \cite{kang2019strong, seo2019strong, bultinck2020ground, lian2020tbg} combine strong interactions and topological bands to predict insulators at all integer fillings. Analytic approaches starting from the chiral flat limit \cite{tarnopolsky2019origin} predict that insulators at integer electron filling $\nu$ are generalized quantum Hall ferromagnets with a quantized anomalous Hall conductance whose parity matches the filling: $\sigma_{xy} = \tfrac{e^2}{\hbar} C$ where $C =\nu\pmod 2$. This prediction holds for the insulators observed at $\nu = -2, 0, 2$ \cite{cao_correlated_2018};  at $\nu = 3$ when samples are aligned with the boron nitride substrate \cite{Sharpe2021_Chern_align, YoungQAH}; and in moderate magnetic fields \cite{Nuckolls2020_Chern_B}. However, a notable exception is found in unaligned samples at $B=0$: most exhibit a $C=0$ insulator at $\nu = 3$ and a metal at $\nu = -3$ \cite{lu_superconductors_2019, Efetov2020Screening, jaoui_quantum_2022, stepanov_competing_2021, cao_correlated_2018, zondiner_cascade_2020, park_flavour_2021, cao2020_nematicity, Yankowitz2019, saito_independent_2020, wong_cascade_2020, oh_evidence_2021, choi2021_STM, yu_correlated_2022} (see App. \ref{app:TBG_experiments}). Thus, at least at $\n{\nu}=3$, generalized QAH ferromagnets must give way to another order, and several candidate $C=0$ insulators have been proposed \cite{kang2020non, Kwan2021, xie2022C2T}. In this work we use accurate density matrix renormalization group (DMRG) \cite{White1992_DMRG, Soejima2020_efficient, kang2020non, Parker2021} calculations at $\nu = -3$ to demonstrate that realistic \textit{heterostrain} qualitatively changes the low-temperature physics in a way that leads to excellent agreement with experiment. In particular, performing large-scale, unbiased calculations that include all spin and valley degrees of freedom, we find that heterostrain stabilizes a spin-polarized $C=0$ ``incommensurate Kekul\'{e} spiral'' order \cite{Kwan2021,Wagner2022} and a ``normal metal'', with important implications for the wider TBG phase diagram. 

Realistic models of TBG fall outside the limit of small dispersion required for strong coupling theory. In particular, experimental samples of TBG are generally found to exhibit  heterostrain \cite{huder2018homoheterostrain,Bi2019,Parker2021,mesple2021strainsurvey,dai2021strainoptical,wang2022strain} (i.e., a difference in strain between the two graphene layers) at the seemingly-insignificant level $\eg = 0.1 - 0.7\%$ \cite{ColumbiaSTM,CaltechSTM,PrincetonSTM,kazmierczak2021strainfield}. 
However, the resulting strain in the moir\'e lattice is enhanced by a factor of the inverse twist angle $\varepsilon_{\text{moir\'e}} \propto \eg/\theta$,
i.e. by two orders of magnitude (see Appendix \ref{app:strain}). Even a tiny strain at the graphene level thus leads to a visible distortion of the moir\'e superlattice [Fig.\,\ref{fig:overview}(a)], as found in STM studies of TBG \cite{ColumbiaSTM,CaltechSTM,PrincetonSTM}. As a result, strain dramatically alters the bandstructure \cite{Bi2019}, increasing the bandwidth of the narrow bands from $\SI{2.5}{\milli\electronvolt}$ to ${\sim}\,\SI{16}{\milli\electronvolt}$ at $\eg = 0.2\% $ and ${\sim}\,\SI{40}{\milli\electronvolt}$ by $\eg = 0.5\%$.  Strain is therefore a significant perturbation that places many TBG samples firmly within the intermediate coupling regime.

The phase diagram of TBG is extremely sensitive to heterostrain. Indeed, a previous DMRG study at $\nu=0$ predicts a phase transition from the strong-coupling Kramers-intervalley coherent insulator to a semimetallic phase at only $\eg \sim 0.2\%$ \cite{Parker2021}, consistent with the experimental finding that gapped and semimetallic phases compete \cite{cao_correlated_2018,PabloSC,Yankowitz2019,cao2020_nematicity,LiVafekscreening,park_flavour_2021,sharpe2019emergent,YoungQAH,lu_superconductors_2019,Efetov2020Screening,wu2021chern}. Away from charge neutrality, a comprehensive self-consistent Hartree-Fock (SCHF) study found that strain drives a transition into an ``incommensurate-Kelul\'e spiral (IKS) order''\cite{Kwan2021,Wagner2022}. At $\n{\nu}=3$, the IKS order is a spin-polarized insulating state that preserves time reversal, but breaks $U(1)_{\mathrm{valley}}$ and --- crucially --- has moir\'e-incommensurate translation breaking.

In this work we establish the phase diagram of the $\n{\nu}=3$ filling of TBG in the presence of strain using unbiased DMRG calculations. We show heterostrain of $\eg=0.05\% - 0.1\%$ drives a transition into an IKS phase with incommensurate translation-breaking [Fig.\,\ref{fig:overview}(d)]. This establishes the presence of IKS order in TBG beyond the mean-field level in a model with all eight electron species. The minute amounts of strain needed to stabilize the IKS order suggest that it is the insulator seen in (hBN-unaligned) samples at $\nu=3$.
\begin{figure*}[t]
    \centering
    \includegraphics[width=\textwidth]{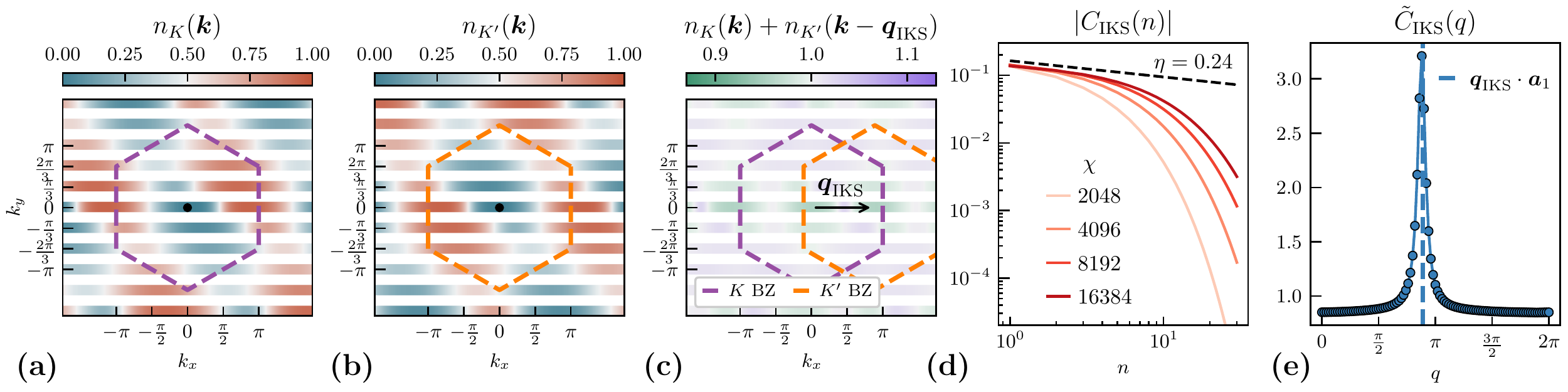}
    \caption{
    a,b) Valley-resolved electron density of TBG at $\nu=-3$, $\eg=0.2 \%$. Dashed hexagons denote the first Brillouin zone, and the dot in the middle is the $\Gamma$ point. c) Total electron density $N_{\qIKS}(\v{k})$ in Eq.\eqref{eq:IKS_invariant_occupations}, after a relative boost by $\qIKS$, whereupon the density becomes uniform.
    d) IKS correlations along the cylinder. As $\chi\to\infty$, correlations approach a power law of Eq.\eqref{eq:IKS_power_law} (black dashed line). e) Fourier transform of the IKS correlation function. The peak at $q \approx 0.89 \pi$ matches c). Parameters: $\kappa =0.65,\ \eg =0.2\%,\ L_y = 6$, spin-polarized \& valley-neutral sector.
    }
    \label{fig:IKS}
\end{figure*}

\PRLsec{Model} We use a standard microscopic model of TBG: eight narrow (``flat") bands from a strained BM model, with strong Coulomb interactions. We overview the Hamiltonian here; see Appendix \ref{app:strain} for details (see also \cite{Soejima2020_efficient,Bi2019,Parker2021}). Sub-percent level heterostrain $\eg$ significantly distorts the moir\'e unit cell [Fig.\,\ref{fig:overview}(a)], leading to lattice vectors $\v{a}_i$ and reciprocal vectors $\v{g}_i$. We use the BM model at twist angle $\theta = 1.08^\circ$, and take chiral ratio $\kappa = w_\mathrm{AA}/w_\mathrm{AB}= 0.5 - 0.8$ to account for some lattice relaxation \cite{nam2017lattice,carr2019exact,ledwith2021tb}. Strain is modelled by adding an effective vector potential to the BM Dirac cones \cite{suzuura2002phonons,manes2007symmetry,kim2008graphene,guinea2008gauge,pereira2009strain,vozmediano2010gauge, de2012space,manes2013generalized,de2013gauge,nam2017lattice,Bi2019,koshino2020effective}. Remarkably, realistic strain increases the bandwidth of the `flat' bands by a factor of $5-10$ relative to $\eg = 0$ [Fig.\,\ref{fig:overview}(b)].

At the many-body level, we use eight species of fermions $\hat{c}^\dagger_{\v{k},\sigma\tau s}$, where
$\sigma =A/B, \tau = K/K', s =\uparrow/\downarrow$ label sublattice \cite{bultinck2020ground}, valley, and spin respectively. The Hamiltonian is (see App.\ref{app:hamiltonian})
\begin{equation}
	\hat{H} = \sum_{\v{k}} \hat{c}_{\v{k}}^\dagger h_{\v{k}} \hat{c}_{\v{k}} + :\frac{1}{2}\sum_{\v{q}} V_{\v{q}} \hat{\rho}_{\v{q}} \hat{\rho}_{-\v{q}}:,
 	\label{eq:interacting_TBG}
\end{equation}
where $\hat{\rho}_{\v{q}}$ is the density at momentum $\v{q}$ and $V_\q$ gives gate-screened Coulomb interactions. As usual, the dispersion $h_{\v{k}}$ is the sum of the BM part and contributions from integrating out the remote bands \cite{bultinck2020ground, Soejima2020_efficient}. Separate charge and spin conservation in each valley give a U$(2) \times$U$(2)$ continuous symmetry (we neglect anisotropies that are expected to enter at the \SI{0.1}{meV} level \cite{bultinck2020ground}). Strain strongly breaks $C_{3z}$ and $C_{2x}$ symmetry, but preserves $C_{2z}$ and time-reversal. The model studied here is very close to particle-hole symmetric \cite{Kwan2021}, and hence DMRG results at $\nu = -3$ and $3$ will be nearly indistinguishable. For notational simplicity we study $\nu = -3$, but our results should not be taken to distinguish between the two. We will conclude by interpreting our results in light of the PH-breaking observed in experiment.

Our DMRG calculations are performed on an infinite cylinder geometry with $L_y$ moir\'e unit cells in the compact direction. We choose a computational `cylinder' basis $\hat{c}_{n, k_y, \sigma\tau s}$ of hybrid Wannier orbitals that are maximally (exponentially) localized at $n$\textsuperscript{th} unit cell along the cylinder axis, but extended around the circumference with definite momentum $k_y$ \cite{Motruk2016_mixedxk,kang2020non,Soejima2020_efficient}. Fourier transformed, our model captures $L_y$ line cuts through the moir\'e Brillouin zone at $k_y = \tfrac{2	\pi m}{L_y}$, $-\tfrac{L_y}{2} \le m < \tfrac{L_y}{2}$. MPO compression \cite{parker2020local, Soejima2020_efficient} is used to faithfully encode the long-range interactions of Eq. \eqref{eq:interacting_TBG} to accuracy $<10^{-2}$ \si{meV} at all distances. We highlight that all eight electron flavors are dynamical in our model. To our knowledge, no other DMRG studies of TBG have included all eight flavors. Our simulations required significant numerical resources. For instance, encoding the $L_y=4$ Hamiltonian requires MPO bond dimension $\chi_{\mathrm{MPO}} \approx 2000$, and we consider states up to $\chi = 24576$. Each unit cell on our cylinder consists of $L_y \times 8$ orbitals, already beyond normal exact diagonalization. Each datapoint requires ${\sim} 40000$ core-hours. By comparison, exact diagonalization studies \cite{xie2021twisted,potasz2021exact} retain at most $3\times3$ unit cells at $\n{\nu}=3$.

\PRLsec{Flavor polarization} Experiments at $\nu=3$ show singly-degenerate quantum oscillations \cite{Nuckolls2020_Chern_B, Yankowitz2019, yu_correlated_2022, lu_superconductors_2019}, indicating flavor symmetry breaking, but the detailed flavor ordering remains elusive.
Our DMRG calculations conserve charge, spin, and valley, allowing us to find the ground state in each quantum number sector. We first focus on the fully spin-polarized sector with neutral valley charge $(\tau^z,s^z) = (0,1)$, where IKS order is present.

\PRLsec{Incommensurate Kekul\'e Spiral}
The IKS is an intervalley coherent (e.g. Kekul\'e) state in which the intervalley $U(1)$ order parameter $\theta$ is modulated in space:  $\theta(\mathbf{r}) \sim \theta_0 + \v{r}\cdot\qIKS$.
IKS order preserves time-reversal, but breaks both $U(1)_\text{valley}$ and moir\'e translation symmetry $\hat{T}_{\v{a}_i}$ down to a combined symmetry 
\begin{equation}
\hat{T}^{\mathrm{IKS}}_{\v{a}_i} = \hat{T}_{\v{a}_i} e^{i \qIKS \cdot \v{a}_i \tau^z/2},
\end{equation}
where $\qIKS$ is \textit{incommensurate} with the moire reciprocal lattice. This results in a state with no charge-density wave at moir\'e scale, but changing Kekul\'e pattern between moir\'e unit cells \cite{Kwan2021, hong2021detecting}.

At $\n{\nu} = 3$, the IKS order additionally breaks spin rotation symmetry and has a non-zero spin polarization. The order parameter manifold of the spin-polarized IKS state thus corresponds to the orbit of the following order parameter under the U$(2)\times$U$(2)$ symmetry action,
\begin{equation}
\hat{\Delta}_\text{IKS}(\qIKS) = \sum_{\v{k}}\hat{c}^\dagger_{\v{k} + \qIKS} P_\uparrow \sigma^x \tau^+ \hat{c}_{\v{k}},\label{eq:IKSordp}
\end{equation}
where $P_\uparrow$ projects on the spin up component. Concretely, the order parameter manifold is given by the space of matrices $U_+^\dagger P_\uparrow U_-$, where $U_{\pm}$ implements the spin/charge symmetry action in valley $\pm K$. This space is SU$(2)\times$SU$(2)/$U$(1)$, where U$(1)$ corresponds to the group of identical spin rotations along the $z$-axis in both valleys.

At zero temperature in two spatial dimensions, the IKS state has true long-range order. However, in the quasi-1D cylinder geometry used in our DMRG simulations, the situation is more subtle. Despite the tendency of strong fluctuations to destroy symmetry breaking in 1+1D \cite{coleman1973there, Hohenberg1967, Mermin1966}, the spin rotation symmetry can be spontaneously broken because the spin polarization order parameter commutes with the Hamiltonian, and hence does not suffer from quantum fluctuations. In the completely spin-polarized sector, the order parameter manifold of the IKS state at $|\nu|=3$ becomes  U$(1)$. This is the same universality class as the 2D XY or 1D XXZ model, and we expect that the spin-polarized IKS state will show up in cylinder DMRG as a phase with algebraic correlations of the IKS order parameter in Eq. \eqref{eq:IKSordp}.

We devised two schemes to identify the quasi-long-range IKS order and the value of $\v{q}_\text{IKS}$ from the ground states on the cylinder: 1) a heuristic ``Brillouin zone shift'' method, 2) finding algebraic correlations of the IKS order parameter.
We first focus on $\eg=0.2 \%$ at filling $\nu=-3$.

The Brillouin zone shift method is based on the Slater determinant representation  of the IKS \cite{Kwan2021}. 
As an insulator, we expect constant electron occupation $n(\v{k})$ in momentum space for such a state. However, due to $\hat{T}^{\mathrm{IKS}}_{\v{a}_i}$ symmetry, we first need to shift the two valleys by $\qIKS$ in order to obtain uniform occupation of the Brillouin zone:
\begin{equation}
    N_{\v{q}_\text{IKS}}(\v{k}) = n_K(\v{k}) + n_{K'}(\v{k} - \v{q}_\text{IKS}) = 1
    	\label{eq:IKS_invariant_occupations}
\end{equation}
To the extent that the true ground state reflects this expectation, the condition $N_{\v{q}_\text{IKS}}(\v{k}) \approx 1$ can be used to infer $\qIKS$.

In Fig.~\ref{fig:IKS}(a, b), we show the electron density $n_{K, K'}(\v{k})$ in each valley of the DMRG ground state, computed by taking a Fourier transform of the electron correlation matrix in the cylinder basis. While electron densities respect time-reversal i.e. $n_K(-\v{k}) = n_{K'}(\v{k})$, the total electron density $n_K(\v{k}) + n_{K'}(\v{k})$ is highly non-uniform. In particular, the occupation in both valleys dips to zero at the $\Gamma$ point, reflecting  the effective band dispersion once accounting for the Hartree interaction with the finite density of $\nu = -3$ of holes \cite{Kwan2021}. However, a shift by $\v{q}_\text{IKS} = 0.448\v{g}_1$ reveals $N_{\qIKS}(\v{k}) \approx 1$ is nearly uniform [Fig.~\ref{fig:IKS} (c)], consistent with the `model' IKS Slater determinant state projected into the $\tau^z=0$ charge sector.

We next investigate the order parameter more directly in the cylinder basis. It is (see App.~\ref{app:DMRG_details} for details)

\begin{equation}
    \hat{\Delta}_{\mathrm{IKS}}(n; q_y) =  \sum_{k_y}\hat{c}^\dagger_{n, k_y + q_y} \sigma^x \tau^+ \hat{c}_{n, k_y}.
	\label{eq:IKS_order_parameter}
\end{equation}
Recall here that $n$ indexes the unit cells along the cylinder.
While the expectation value of this operator is always zero due to $U(1)_\text{valley}$ conservation, its correlator $C_\text{IKS}(n;q_y) = \langle \hat\Delta_\text{IKS}(n;q_y)\hat\Delta^\dagger_\text{IKS}(0;q_y)\rangle$ can show algebraic correlation (see App.\,\ref{app:DMRG_details})
\begin{equation}
    \CIKS(n \gg 1; q_y = \qIKSy) \propto n^{-\eta} e^{i(\qIKS\cdot \v{a}_1)n},
	\label{eq:IKS_power_law}
\end{equation}
where the phase factor reflects the translation-breaking nature of the IKS, and algebraic decay is only observed at $q_y=\qIKSy$ (see App.\,\ref{app:DMRG_details}).  Due to the finite DMRG bond dimension $\chi$, the correlations will decay at long distance as $\CIKS \sim e^{-n/\xi_\mathrm{IKS}(\chi)}$, with Eq.~\eqref{eq:IKS_power_law} recovered only in the limit $\chi \to \infty$. In Fig.~\ref{fig:IKS}(d) we show that the correlations indeed approaches a power law for a particular choice of $q_y$; the resulting exponent can be calculated via ``finite entanglement scaling'' \cite{pollmann2009_finite_entanglement, tagliacozzo2008scaling, pirvu2012matrix}, see App.\,\ref{app:IKS}.

Finally, the correlator gives $\v{q}_\textrm{IKS}$ directly. First, we can determine $q_\mathrm{IKS}^y=0$ since it gives the largest correlation length in $\CIKS(n;q_y)$. Moreover, the discrete Fourier transform of such correlator with respect to $n$, denoted as $\tilde{C}_\textrm{IKS}(q)$, reveals a peak at $\qIKS \cdot \v{a}_1/(2\pi) \approx 0.445$ (Fig.~\ref{fig:IKS} (e)). Combining these gives $\v{q}_\text{IKS} = 0.445\v{g}_1$ , consistent with that found from Eq.~\eqref{eq:IKS_invariant_occupations}.

Summing up, we have shown the  spin-polarized, valley-neutral ground state at $\nu=-3$ is consistent with a 2D phase that breaks $U(1)_\text{valley}$ and translation symmetry but preserves $\hat{T}_{\v{a}_i}^{\mathrm{IKS}}$ and time-reversal symmetry, which are the defining properties of the IKS order. We have checked that this order is robust to changing: the chiral ratio $\kappa$ [Fig.\,\ref{fig:overview} (d)], cylinder circumference, the strength and direction of heterostrain, and interaction strength  (but not to hBN alignment \cite{Kwan2021}). IKS order is therefore remarkably flexible and robust. 

\PRLsec{Strain favors $\Gamma$-depleted states}
Why is IKS order favored in the intermediate coupling regime? A key reason is electron-depletion near the $\Gamma$ point (Fig.~\ref{fig:IKS}) \cite{Kwan2021}. A combination of interaction effects and the heterostrain-driven dispersion gives rise to an energy peak near the $\Gamma$ point (App. \ref{app:hamiltonian}).
As a momentum-offset superposition between different valley flavors, IKS order evades populating this region while still avoiding the exchange penalty resulting from a Fermi surface. In the absence of spin polarization, there are more ways to avoid populating the $\Gamma$ point, giving rise to a set of $\Gamma$-depleted states with symmetry breaking. If heterostrain is large enough, on the other hand, we expect a metallic state due to the large single-particle dispersion.
We now confirm these expectations by investigating, as a function of heterostrain, four quantum number sectors where valley and spin are either polarized or neutral: $(\tau^z,s^z) = (1,1), (1,0), (0,1), (0,0)$. See App.\,\ref{app:DMRG_details} for precise details of symmetry sectors.

\renewcommand{\arraystretch}{1.4}
\begin{table}[t]
    \begin{tabular}{lcccccclc}
    \toprule Phase & $\tau^z$ & $s^z$ & $U(1)_{V}$ & $\mathcal{T}$ &  Translation  &   $\n{C}$\\ \hline
    QAH & 1 & 1 & $\checkmark$ & $\times$ &  $\hat{T}_{\v{a}_i}$ & 1\\
    QAH-IVC & 0 & 1 & $\checkmark$ & $\times$ & $\hat{T}_{\v{a}_i}$ & 1\\
    \hline
    NSM & 1 & 1 & $\checkmark$ & $\times$ & $\hat{T}_{\v{a}_{i}}$ &  0 \\
    IKS & 0 & 1 & $\times$ & $\checkmark$ & $\hat{T}^{\mathrm{IKS}}_{\v{a}_i} = \hat{T}_{\v{a}_{i}} e^{i \v{q} \cdot \v{a}_{i} \tau_{z} / 2}$ & 0 \\
    CSS & 1 & 0 & $\checkmark$ & $\times$ & $\hat{T}^{\mathrm{CSS}}_{\v{a}_i} = \hat{T}_{\v{a}_{i}} e^{i \v{q}' \cdot \v{a}_{i} s_{z} / 2}$ & 0 \\
    NM & 0 & 0 & $\checkmark$ & $\checkmark$ & $\hat{T}_{\v{a}_{i}}$ &  0 \\
    \botrule
    \end{tabular}
    \caption{Ground state candidates at $\n{\nu} =3$. Here $\tau^z$ and $s^z$ specify the flavor polarization, $U(1)_V$ is valley conservation, $\mathcal{T}$ is time-reversal, and $C$ is the Chern number. The top and bottom sections are energetically competitive at strain $\eg =0$ and $\eg \ge 0.05\%$, respectively.
    (QAH) flavor-polarized quantum anomalous Hall; (QAH-IVC) spin-polarized, intervalley coherent QAH; (NSM) flavor-polarized nematic semimetal; (IKS) incommensurate kekule spiral; (CSS) commensurate spin-spiral; (NM) fully symmetric ``normal'' metal. 
    }
    \label{tb:gs_phase_candidates}
\end{table}

\begin{figure}
    \centering
    \includegraphics[width=\linewidth]{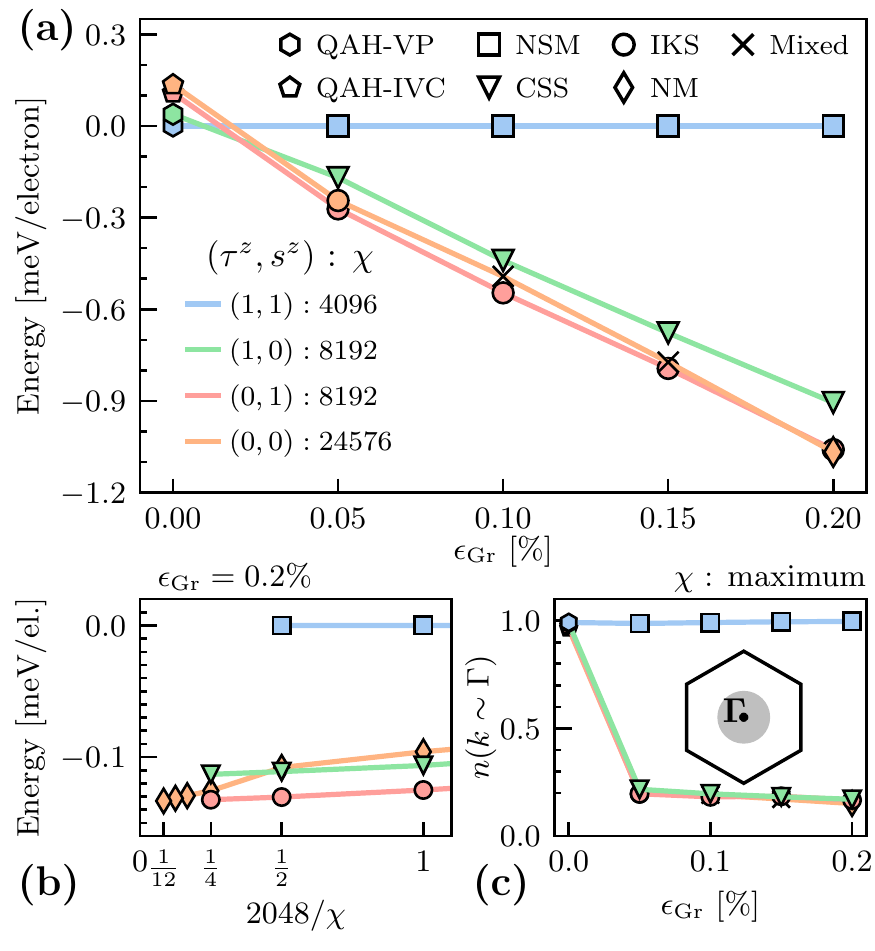}
    \caption{a) Ground state energy as a function of heterostrain $\eg$ for various flavor polarizations. Energy is measured relative to the spin- and valley-polarized ground state, and symbols correspond to phases in Table \ref{tb:gs_phase_candidates}. Heterostrain of only $\eg \approx 0.025\%$ drives a transition from QAH-type order to IKS-type order. b) Ground state competition between flavor polarization sectors at $\eg = 0.2\%$. c) Electron density per momentum near $\Gamma$: $n(k \sim\Gamma) = \sum_{\n{k} < 1/5} n(k)$. 
	Parameters match Fig.\,\ref{fig:IKS}, except with $L_y = 4$.
	}
    \label{fig:phase_diagram}
\end{figure}

In Fig.\,\ref{fig:phase_diagram}(c), we show the electron density near the $\Gamma$ point for different quantum number sectors. Above a low heterostrain of $\epsilon_{Gr} = 0.05 \%$, all but the fully flavor polarized sector has a substantial reduction in $\Gamma$ electron population.

At $\eg = 0$, we find the ground state is a spin-polarized QAH insulator consistent with strong-coupling theory. This phase  is detected via the $\mathcal{T}$-breaking order parameter $\hat{\Delta}_{\rm{QAH}} = \sum_{\v{k}} \hat{c}^\dagger_{\v{k}} \sigma^z \tau^z \hat{c}_{\v{k}}$ (see App.\,\ref{app:phase_identification} for further details.)
When $s^z=0$, we find long-range spin correlations consistent with spin polarization into the $xy$-plane.
The QAH state comes in two nearly-degenerate varieties: a valley-polarized QAH when $\tau^z=1$, and an inter-valley coherent QAH when $\tau^z = 0$ [Fig.\,\ref{fig:phase_diagram}(a)]. At $\kappa=0$, an emergent U$(4)_+ \times$U$(4)_-$ symmetry rotates these states into each other \cite{bultinck2020ground}. Since the QAH and QAH-IVC states are physically similar, it is unsurprising that they remain nearly degenerate at $\kappa=0.65$. Therefore the predictions of strong coupling theory are borne out at $\eg=0$.

The phases found at $\eg \geq 0.05 \%$ are summarized in Table \ref{tb:gs_phase_candidates}, and App.\,\ref{app:phase_identification}-\ref{app:neutral_sector} give numerical details of their identification. The ground state in the spin- and valley-polarized sector, $(1,1)$, is a nematic semimetal (NSM) with two Dirac cones near  the $\Gamma$-point in each valley \cite{liu2021_NSM,Soejima2020_efficient,Parker2021,Kwan2021}. This state is at relatively high energy, as it cannot benefit from $\Gamma$-depletion. The spin-neutral valley-polarized sector $(1,0)$ has a commensurate spin-spiral (CSS) order, characterized by the order parameter $\hat{\Delta}_{\textrm{CSS}} = \hat{c}^\dagger_{\v{k}+\v{q}'} s^x \hat{c}_{\v{k}} + h.c.$ where $\v{q}' = \frac{\v{g}_2}{2}$. In the 2D limit, this doubles the unit cell, but respects a combined translation-times-spin-rotation symmetry  $\hat{T}^{\mathrm{CSS}}_{\v{a}_i}$. The $\mathbf{q'}=\mathbf{g}_2/2$ CSS is collinear, and thus has an unbroken U$(1)$ spin rotation symmetry. In the valley-neutral, spin-polarized sector $(0,1)$, IKS order \eqref{eq:IKS_order_parameter} is the ground state, as previously discussed. 

Finally, the fully-neutral sector $(0,0)$ features both flavor-polarized and unpolarized states: At $\eg = 0.05\%$, we find IKS order with spontaneous spin-polarization into the $xy$-plane. Both the energy and $\qIKS$ vector match the IKS order in the $(0,1)$ sector, suggesting that spin-polarized IKS is the ground state in the moderate strain regime. At higher strain $\eg = 0.2\%$ we find a state consistent with a fully-symmetric  ``normal metal" (NM).
This state has four nascent flavor-degenerate Fermi surfaces which sharpen with the DMRG bond dimension and closely match the flavor-symmetric metal found within Hartree-Fock (App.\,\ref{app:symmetric_metal}).
Since the resulting metal should have central charge $c = 4L_y = 16$, it is exceptionally difficult to converge with DMRG, and we are not able to perform a definitive scaling analysis even at $\chi=24576$.
For $\eg = 0.1-0.15\%$, the same signatures of the putative ``normal metal" are present, but coexist with a strong spin-density wave and valley/spin fluctuations. This region, close to the transition between flavor polarized and unpolarized phases at intermediate strain, is denoted as ``mixed" order (see App. \ref{app:mixed_order}).

Fig.\,\ref{fig:phase_diagram}(b) shows the remarkably close energetic competition between the flavor polarization sectors at $\epsilon_{Gr}=0.2\%$. This suggests there are competing orders which are split at ${\sim}\,\SI{0.1}{\milli\electronvolt}$, two orders of magnitude below the Coulomb scale --- a phenomenon found in other intermediate coupling models (see e.g. \cite{arovas2022hubbard}). Note that this energy difference is much smaller than the uncertainty in the Hamiltonian itself. So while we may conclude the high-strain ground state is likely to live somewhere in the set of $\Gamma$-depleting states, our present numerics do not definitively choose a unique order.

\PRLsec{Experimental Implications} We briefly conclude with experimental implications. Our results suggest that the amounts of heterostrain expected to be present in virtually all experimental samples are more than sufficient to push TBG into the intermediate coupling regime. In this regime, our numerics suggest IKS order is the primary insulating ground state candidate at $\n{\nu} = 3$. We note this is specific to samples \textit{not} aligned to an hBN substrate; alignment strongly favors the QAH phase. Although we have used a realistic microscopic model \eqref{eq:interacting_TBG}, there are a number of phenomena on the \si{\milli\electronvolt} scale we have not captured, such as for example lattice relaxation and particle-hole symmetry breaking   \cite{kang2022_PHbreaking}. The great majority of experiments (Table in App.~\ref{app:TBG_experiments}) find a $C=0$ insulator at $\nu=3$ --- consistent with IKS --- but a metallic state at $\nu=-3$.
We suggest the metallic state might be the ``normal metal" we find at $\eg = 0.2\%$ and above. Our prediction of an IKS phase at $\nu = 3$ could be directly confirmed by graphene-scale STM experiments \cite{hong2021detecting,calugaru2022KekuleSTMtheory}.

\begin{acknowledgments}

We thank Patrick Ledwith, Ilya Esterlis, Eslam Khalaf, Ashvin Vishwanath, Yves Kwan, Glenn Wagner, Steve Simon and Siddharth Parameswaran for insightful discussions and collaborations on related works. This research used resources of the National Energy Research Scientific Computing Center (NERSC), a U.S. Department of Energy Office of Science User Facility located at Lawrence Berkeley National Laboratory, operated under Contract No. DE-AC02-05CH11231 using NERSC award BES-ERCAP0020043. 
This work was supported by the U.S. Department of Energy, Office of Science, Basic Energy Sciences, under Early Career Award No. DE-SC0022716, and the Alfred P. Sloan Foundation.
This research is funded in part by the Gordon and Betty
Moore Foundation’s EPiQS Initiative, Grant GBMF8683
to D.E.P. T.S. is supported by Masason foundation. N.B. is supported by a University Research Fellowship of the Royal Society. This research used the Lawrencium computational cluster resource provided by the IT Division at the Lawrence Berkeley National Laboratory (Supported by the Director, Office of Science, Office of Basic Energy Sciences, of the U.S. Department of Energy under Contract No. DE-AC02-05CH11231).

\end{acknowledgments}

\bibliography{bibliography,bibliography_review}

\pagebreak

\newpage

\onecolumngrid
\begin{center}

\textbf{\large Supplementary Information}
\end{center}
\setcounter{equation}{0}
\setcounter{figure}{0}
\setcounter{table}{0}
\makeatletter
\renewcommand{\thefigure}{S\arabic{figure}}
\renewcommand{\thesection}{S\Roman{section}}
\renewcommand{\bibnumfmt}[1]{[S#1]}
\setcounter{section}{0}

\appendix

The Appendices to this work are organized as follows.

\begin{description}
    \item[Appendix \ref{app:TBG_experiments}] overviews insulators found in experiments at integer fillings of TBG.
    \item[Appendix \ref{app:strain}] shows how small strains on the graphene scale are amplified at the moir\'e scale.
    \item[Appendix \ref{app:hamiltonian}] is a self-contained description of the Hamiltonian and computational basis used in this work. This section also describes the ``Hartree peak".
    \item[Appendix \ref{app:DMRG_details}] gives numerical details of our DMRG computations, including the quantum number sectors used in the main text, and defines some common correlation functions.
    \item[Appendix \ref{app:phase_identification}] details the phases that appear at $\nu=-3$ \textit{not} in the $(0,0)$ sector and discusses their identification in the thin-cylinder limit. 
    \begin{description}
        \item[\ref{app:QAH}] the quantized anomalous Hall (QAH) phase, and its intervalley coherent variant (QAH-IVC).
        \item[\ref{app:CNSM}]  the compensated semimetal in the $(1,1)$ sector.
        \item[\ref{app:CSS}]  the commensurate spin spiral (CSS) phase.
        \item[\ref{app:IKS}]  further details of the IKS order in the spin-polarized $(0,1)$ sector.
    \end{description}
    \item[Appendix \ref{app:heisenburg_analogy}] is a brief analogy to the 1D Heisenberg model in the $s^z =0$ sector, which highlights some features that will appear in the $(0,0)$ sector.
    \item[Appendix \ref{app:neutral_sector}] details the phases that appear in the $(0,0)$ sector.
    \begin{description}
        \item[\ref{app:symmetric_metal}]  the symmetric metallic order at large strain in the $(0,0)$ sector, both in DMRG and its manifestation within self-consistent Hartree-Fock.
        \item[\ref{app:spinfulIKS}]  the spinful IKS order in the $(0,0)$ sector at $\eg = 0.05\%$.
        \item[\ref{app:mixed_order}]  aspects of the ``mixed" order at intermediate strain in the $(0,0)$ sector.
        \item[\ref{app:overall_phase_diagram}] The putative phase diagram of $\nu=-3$ as a function of $\eg$, including higher-strain data.
    \end{description}
\end{description}

\section{Overview of TBG Experiments at integer fillings}
\label{app:TBG_experiments}

This appendix reviews measurements of integer filling states in TBG at zero field and without hBN-alignment. We focus on correlated insulating (CoI) states, and consider three broad classes of experimental probes:
\begin{description}
    \item[Transport measurements] Direct measurements of resistance $\rho_{xx}$. Correlated insulators manifest as a peak in resistance. As these measurements go across the entire sample, they are  affected by sample inhomogeneity (strain, twist angle disorder, etc). This tends to suppress the signals associated with insulating states relative to local probes.
    \item[STM] By measuring the local density of states above a state at a given filling as a function of bias voltage, STM can directly measure the one-electron gaps. Some experiments use point-contact spectroscopy to distinguish superconductors (SC) from correlated insulators.
    \item[SET] Single electron transistors are a local probe that directly measures the inverse compressibility $\frac{d\mu}{dn}$. Insulating states appear as peaks in $\frac{d\mu}{dn}$, which may be integrated to estimate the spectral gap. SET measurements can be used to infer where flavor polarization occurs.
\end{description}

Table \ref{tab:experiment_review} gives a non-exhaustive overview of experimental results at all integer fillings. As discussed in the introduction, the results display a strong particle-hole breaking pattern, especially at filling $\nu=\pm3$.

\begin{table}[]
    \centering
    \begin{ruledtabular}
        \begin{tabular}{l|ccc|ccccccc|c}
Report & $\theta$ (\textdegree) & Gate type & $D$ (nm) & $-3$ & $-2$ & $-1$ & $0$ & $1$ & $2$ & $3$ & Notes \\
\hline
\textit{\textbf{Transport}} &&&&&&&&&&&\\
Ref.\,\cite{lu_superconductors_2019} & 1.10 & single & $10$ & (CoI) & CoI & (CoI) & CoI & CoI & CoI & CoI & \SI{16}{\milli\kelvin} \\
Ref.\,\cite{Efetov2020Screening}, D1 & 1.15 & single & 7 & & & & CoI & & & & \SI{25}{\milli\kelvin}\\
Ref.\,\cite{Efetov2020Screening}, D2 & 1.04 & single & 9.8 & & & & CoI & & (CoI) & & \SI{25}{\milli\kelvin}\\
Ref.\,\cite{Efetov2020Screening}, D3 & 1.10 & single & 12.5 & & CoI & & CoI & (CoI) & CoI & CoI & \SI{25}{\milli\kelvin}\\
Ref.\,\cite{jaoui_quantum_2022} & 1.04 & single & 9.5 & SC & & & CoI & & (SC) & CoI & \SI{40}{\milli\kelvin} \\
Ref.\,\cite{stepanov_competing_2021} & 1.08 & single & 7 & & CoI & & CoI & ``ChI" & CoI & CoI & \SI{30}{\milli\kelvin} \\
Ref.\,\cite{cao_correlated_2018}, D1 & 1.08 & single & 10-30 & & CoI & & CoI & & CoI &  & \SI{300}{\milli\kelvin} \\
Ref.\,\cite{cao_correlated_2018}, D2 & 1.10 & single & 10-30 & & (CoI) & & CoI & & CoI & & \SI{300}{\milli\kelvin} \\
Ref.\,\cite{cao_correlated_2018}, D3 & 1.12 & single & 10-30 & & CoI & & & & CoI & (CoI) & \SI{100}{\milli\kelvin} \\
Ref.\,\cite{cao_correlated_2018}, D4 & 1.16 & single & 10-30 & & CoI & & CoI & & CoI & (CoI) & \SI{300}{\milli\kelvin} \\
Ref.\,\cite{zondiner_cascade_2020}& $\sim$1.07 & single & 42 & & CoI & & CoI & & CoI & CoI & \SI{50}{\milli\kelvin}\\
Ref.\,\cite{park_flavour_2021} & 1.07 & dual & ? & & CoI & & CoI & (CoI) & CoI & CoI & \SI{70}{\milli\kelvin}\\
Ref.\,\cite{cao2020_nematicity}, DA & 1.09 & single & 50 & & CoI & & CoI & (CoI) & CoI & CoI & \SI{70}{\milli\kelvin}\\
Ref.\,\cite{Yankowitz2019}, D1 & 1.14 & dual & 30-60 & & (CoI) & & CoI & & CoI & CoI & \SI{10}{\milli\kelvin}\\
Ref.\,\cite{Yankowitz2019}, D3 & 1.10 & dual & 30-60 & CoI & CoI & & CoI & & CoI & CoI & \SI{300}{\milli\kelvin}\\
Ref.\,\cite{Yankowitz2019}, D5 & 1.08 & dual & 30-60 & & CoI & & CoI & (CoI) & CoI & CoI & \SI{10}{\milli\kelvin}\\
Ref.\,\cite{saito_independent_2020}, D1 & 1.08 & single & 68 & & CoI & & CoI & (CoI) & CoI & CoI & \SI{50}{\milli\kelvin}\\
Ref.\,\cite{saito_independent_2020}, D2 & 1.09 & single & 6.7 & & CoI & & CoI & & CoI & CoI & \SI{50}{\milli\kelvin}\\
Ref.\,\cite{saito_independent_2020}, D3 & 1.04 & single & 38 & & & & CoI & & CoI & CoI & \SI{50}{\milli\kelvin}\\
Ref.\,\cite{saito_independent_2020}, D4 & 1.18 & single & 7.5 & & & & CoI & & & & \SI{50}{\milli\kelvin}\\
Ref.\,\cite{saito_independent_2020}, D5 (\cite{saito_hofstadter_2021}) & 1.12 & single & 45 (40) & & CoI & & CoI & & CoI & CoI & \SI{10}{\milli\kelvin} \\
\hline
\textit{\textbf{STM}} &&&&&&&&&&&\\
Ref.\,\cite{wong_cascade_2020}(\cite{oh_evidence_2021}, DB) & 1.06 & STM tip & - & (SC) & CoI & CoI & CoI & CoI & CoI & CoI & $\eg=0.1\%$ \\
Ref.\,\cite{oh_evidence_2021}, DA & 1.13 & STM tip & - & SC & CoI & & CoI & & CoI & CoI & $\eg=0.4\%$ \\
Ref.\,\cite{oh_evidence_2021}, DA' & 1.01 & STM tip & - & SC & CoI & & CoI & & CoI & CoI & $\eg=0.2\%$ \\
Ref.\,\cite{choi2021_STM} & $1.27\to 0.97$ & STM tip & - & & CoI & & CoI & (CoI) & CoI & CoI & $\eg<0.3\%$ \\
Ref.\,\cite{choi2021_STM} (Supp.) & $1.26\to0.99$ & STM tip & - & & CoI & & CoI & CoI & CoI & CoI & \\
\hline
\textit{\textbf{SET}} &&&&&&&&&&&\\
Ref.\,\cite{zondiner_cascade_2020}& 1.13 & single & 42 & (CoI) & (CoI) & (CoI) & - & CoI & CoI & CoI & \SI{4}{\kelvin}\\
Ref.\,\cite{yu_correlated_2022} & 1.06 & single & 40 & & (CoI) & & CoI & - & CoI & CoI & \SI{330}{\milli\kelvin}\\
        \end{tabular}
    \end{ruledtabular}
    \caption{An overview of integer filling states in TBG visible in selected experimental data. This list is non-exhaustive and is restricted to samples not aligned to hBN and without an applied magnetic field.     Correlated insulators (CoI), superconductors (SC), and Chern insulators (ChI), are observed at the labeled integer fillings. If an insulator at filling $+\nu$ is substantially weaker than those at other fillings, in particular its particle-hole partner at $-\nu$, then it is shown in parentheses, e.g. as ``(CoI)". We have labelled integer filling states as ``SC" if there is a superconductor \textit{near} that integer filling, e.g. at $\nu=-3+0.1$.
    }
    \label{tab:experiment_review}
\end{table}

\section{Effect of Strain on the Moir\'e Superlattice}
\label{app:strain}

This appendix shows how heterostrain acts on the moir\'e superlattice of twisted bilayer graphene. We will conclude that its effect is enhanced by a factor of the inverse of twist angle.

Consider two layers of graphene ($\ell=\pm$) with opposite twist angle $\theta_\ell = \ell\theta/2$ and uniaxial strain $\varepsilon_\ell = \ell\varepsilon/2$. Suppose $\v{R}_i$ are the lattice vectors of graphene and $\v{G}_j$ are the corresponding reciprocal lattice vectors so that $\v{R}_i \cdot \v{G}_j = 2\pi \delta_{ij}$. The twisted lattices on each layer are then given by
\begin{equation}
    \v{R}^{\ell}_{i} = R(\theta_\ell)S(\varepsilon_\ell)\v{R}_{i},\quad i=1,2,
\end{equation}
where $R,S$ are matrix representations of rotation and strain operation:
\begin{equation}
    R(\theta) = \begin{pmatrix} 
      \cos{\theta}     & -\sin{\theta}\\ 
      \sin{\theta} & \cos{\theta}
    \end{pmatrix},\quad
    S(\varepsilon) = \begin{pmatrix} 
      1+\varepsilon     & 0\\ 
      0 & 1-\nu\varepsilon
    \end{pmatrix},
\end{equation}
where $\nu=0.16$ is the Poisson ratio of graphene.

For $\theta, \varepsilon \ll 1$, this is approximated to leading order by
\begin{equation}
    \v{R}^{\ell}_{i} \approx M(\theta_\ell,\varepsilon_\ell)\v{R}_{i}; 
    \quad M(\theta, \varepsilon) = \begin{pmatrix} 
      1+\varepsilon & -\theta\\ 
      \theta & 1-\nu\varepsilon
    \end{pmatrix}.
    \label{eq:strain1}
\end{equation}
Similarly, the reciprocal lattice vectors of each layer are
\begin{equation}
    \v{G}^{\ell}_{i} \approx \left[M^{-1}(\theta_\ell,-\varepsilon_\ell)\right]^T\v{G}_{i} = M(\theta_\ell,-\varepsilon_\ell)\v{G}_{i}.
\end{equation}

The moir\;e superlattice of the two layers has a reciprocal lattice
\begin{align}
    \v{g}_{i} &= \v{G}^{+}_{i} - \v{G}^{-}_{i}
    = \left(M(\theta/2,-\varepsilon/2) - M(-\theta/2,\varepsilon/2)\right)\v{G}_{i}
    = \begin{pmatrix} 
      -\varepsilon & -\theta\\ 
      \theta & \nu\varepsilon
    \end{pmatrix}\v{G}_{i}.
\end{align}
Note that we can view the contribution of strain separately from the twist operation: 
\begin{equation}
    \v{g}_{i}[\theta,\varepsilon] = \underbrace{\begin{pmatrix} 
      1 & -\varepsilon/\theta \\ 
      -\nu\varepsilon/\theta & 1
    \end{pmatrix}}_{\tilde{M}_\theta(\varepsilon)}
    \begin{pmatrix} 
      0 & -\theta\\ 
      \theta & 0
    \end{pmatrix}\v{G}_{i}
    =\tilde{M}_\theta(\varepsilon)\v{g}_{i}[\theta,0].
\end{equation}
Diagonalizing $\tilde{T}_\theta(\varepsilon)$, we can consider the effect of strain as a dilation on the moir\'e scale:
\begin{equation}
\tilde{S}_\theta(\varepsilon)=  U\tilde{T}_\theta(\varepsilon)U^\dagger = \begin{pmatrix} 
      1 + \sqrt{\nu}\varepsilon/\theta & 0 \\ 
      0& 1 - \sqrt{\nu}\varepsilon/\theta
    \end{pmatrix},
    \label{eq:strain2}
\end{equation}
which acts on transformed reciprocal lattice vectors $\{\tilde{\v{g}}_i = U\v{g}_{i}\}$ and the corresponding superlattice basis $\{\tilde{\v{r}}_i\}$.
\begin{align}
    \tilde{\v{g}}_i [\theta,\varepsilon] &= \tilde{S}_\theta(\varepsilon)\tilde{\v{g}}_i [\theta,0],\\
    \tilde{\v{r}}_i [\theta,\varepsilon] &= \tilde{S}^{-1}_\theta(\varepsilon)\tilde{\v{r}}_i [\theta,0].
\end{align}
Comparing Eq. \eqref{eq:strain1} and \eqref{eq:strain2}, we conclude that heterostrain $\eg$ on graphene lattice is equivalent to an amplified strain 
\begin{equation}
\varepsilon_{\text{moir\'e}} \propto \frac{\eg}{\theta}
\label{eq:strain_amplification}
\end{equation} 
on the moire superlattice. In the experimentally relevant case of  $\theta\sim1$\textdegree, $\eg\sim 0.2\%$, we have $\varepsilon_{\text{moir\'e}}\sim 10\%$, which is not a negligible perturbation.

\section{Microscopic Model of Strained TBG}
\label{app:hamiltonian}

This appendix describes the microscopic model of strained TBG used in the main text. This model is standard; it is essentially identical to the models used in Ref. \cite{Soejima2020_efficient}, which introduced our DMRG method, and Refs. \cite{Bi2019, Parker2021} that studied strained TBG.

\subsection{Single Particle Hamiltonian}

Strained TBG is described at the single-particle level by a generalized Bistritzer-MacDonald Hamiltonian \cite{dos2007graphene,bistritzer2011moire,nam2017lattice,Bi2019}. Strain affects the Hamiltonian in two separate ways. First, as shown above, it distorts the lattice and superlattice. Second, it changes the inter-atomic distance between carbon atoms, thereby modulating the hopping integrals and changing the energetics directly. We incorporate both effects.

Within each layer $\ell = \pm$, we use the standard model of graphene: 
\begin{equation}
    h_{\rm{Gr}}(\v{k}) = -t \begin{pmatrix}
        0 & D(\v{k})\\
        \overline{D(\v{k})} & 0
    \end{pmatrix},\quad 
    D(\v{k}) = \sum_{\mu=0}^2 e^{-i\v{k}\cdot \v{\tau}_\mu},
\end{equation}
where $\v{\tau}_\mu$ are the three vectors from $A$ to $B$ sublattices. We use $t = \SI{2.8}{meV}$ and lattice constant $a_{\rm{Gr}} = \SI{0.236}{\nano\meter}$. At low energies, this model reduces to the Dirac equation $h(\v{k})_{\tau = K} = v_F \v{\sigma}\cdot (-i \hbar \nabla)$. A long line of work (see e.g. \cite{suzuura2002phonons,manes2007symmetry,kim2008graphene,guinea2008gauge,pereira2009strain,vozmediano2010gauge, de2012space,manes2013generalized,de2013gauge,nam2017lattice,Bi2019,koshino2020effective}) has shown that strain on the graphene scale is incorporated into the Dirac equation as an effective vector potential $-i \v{\nabla} \to -i \v{\nabla} + \v{A}$, or $\v{k} \to \v{k} + \v{A}$ in momentum space. For the $K$-valley, we have
\begin{equation}
    h_{\ell = \pm,\tau=K}(\v{k}) = h_{\rm{Gr}}(M(\theta_\ell, -\epsilon_{\ell}) [\v{k} + \v{A}_{\ell,\tau}]),  \quad 
    \v{A}_{\ell} = -\frac{\ell}{2} \frac{\beta \sqrt{3}}{2a} \begin{pmatrix}
        [1-\nu_P] \epsilon, 0 
    \end{pmatrix}
\end{equation}
where $M(\theta,\epsilon)$ is given in Eq. \eqref{eq:strain1},  $\sigma \in \{A,B\}$ label the graphene sublattices, $a = \n{\v{R}_i}$ is the lattice constant of graphene, $\beta \approx 3.14$ characterizes how much the carbon-carbon hopping integral changes under lattice deformations, and $\nu_P \approx 0.16$ is the Poisson ratio of graphene. Crucially $C_{3z}$ symmetry is broken by the lattice distortion. so the Dirac points are no longer pinned to the $K,K'$ points and are shifted by $\v{A}_{\ell}$ \cite{Bi2019}.

We now consider the mori\'e superlattice with reciprocal lattice vectors $\v{g}_i$ defined above. We choose conventions for graphene so that $\v{g}_1$ is along the $x$-axis of reciprocal space (up to corrections from strain). In this setting, the BM model becomes ($K$-valley)
\begin{equation}
	 h_K = \begin{bmatrix} 
	 h_{+,K} & T(\v{r})\\
	 T(\v{r})^\dagger & h_{-,L}
	 \end{bmatrix}, \quad 
	 T(\v{r}) = T_0 + T_1 e^{-i\v{g}_2 \cdot \v{r}} + T_2 e^{i (\v{g}_1 - \v{g}_2) \cdot \v{r}},
	 \quad T_n = w_{AB} \begin{pmatrix}
	     \kappa & e^{-i 2\pi n/3}\\
	     e^{i 2\pi n/3} & \kappa
	 \end{pmatrix}
	 \label{eq:single_particle_strained_TBG}
 \end{equation}
where $T(\v{r})$ is the interlayer tunneling, whose form we assume is unchanged apart from the lattice distortion, with $w_{AB} = \SI{110}{meV}$. Due to lattice relaxation effects, the realistic range for the chiral ratio $\kappa$ is thought to be $\kappa = 0.5 - 0.8$ \cite{nam2017lattice,carr2019exact,ledwith2021tb}. As usual, the harmonics that appear in Eq. \eqref{eq:single_particle_strained_TBG} are the lowest harmonics that connect the $K$-points of the two graphene layers.

The Hamiltonian for both valleys is
\begin{equation}
    h_{\mathrm{TBG}} = \begin{pmatrix}
        h_K \otimes s^0 & 0\\
        0 & h_{K'} \otimes s^0
    \end{pmatrix}
    \label{eq:h_TBG_single_particle}
\end{equation}
where $s^\mu$ are the spin Pauli matrices and $h_{K'}$ is defined as the time-reversal conjugate of $h_{K}$, making $h_{\mathrm{TBG}}$ time-reversal symmetric. In fact, $h_{\mathrm{TBG}}$ inherits the $C_{2z}$ symmetry of graphene. However, both $C_{3z}$ and $C_{2x}$ symmetries are explicitly broken by strain \cite{Bi2019}. If one makes the approximation $T(\theta_\ell, -\epsilon_{\ell}) \to T(0, -\epsilon_{\ell})$ in the Dirac equation, then the single-particle Hamiltonian becomes particle-hole symmetric in the absence of strain \cite{bultinck2020ground} Even in the presence of strain, the Hamiltonian is still largely particle-hole symmetric. We therefore expect the physics at $\pm \nu$ to be at least qualitatively similar (though this is not the case in experiments). 

\subsection{Many-Body Hamiltonian}
For our many-body Hamiltonian, we take the single-particle model discussed above, add gate-screened Coulomb interactions, and integrate out the remote bands at the mean-field level. We review this procedure here; in-depth treatments are given in Refs. \cite{Soejima2020_efficient, bultinck2020ground, hofmann2022fermionic, parker2021field}. Consider a basis $\hat{\v{f}}_{\v{k}}^\dagger = \hat{f}^\dagger_{\v{k} b \tau s}$, where $b$ labels bands of the BM model, $\tau \in \{K,K'\}$ labels valleys, and $s \in \{ \uparrow,\downarrow\}$ indexes spin. We start from the microscopic Hamiltonian
\begin{equation}
    \hat{H}_{\mathrm{full}} 
    = \sum_{\v{k} \in \mathrm{BZ}} \hat{\v{f}}^\dagger_{\v{k}} [ h_{\mathrm{TBG}}(\v{k}) - h_{\mathrm{counter}}(\v{k}) ] \hat{\v{f}}_{\v{k}}
    + \frac{1}{2A} \sum_{\v{q} \in \mathbb{R}^2} V_{\v{q}} : \hat{\v{\rho}}_{\v{q}} \hat{\v{\rho}}_{-\v{q}}:, \quad 
    V_{\v{q}} = \frac{1}{2\epsilon_r \epsilon_0 \v{q}} \tanh(\n{\v{q}} d)
\end{equation}
where the sums on $\v{k}$ always run over the Brillouin zone, while $\v{q}$ is unrestricted, $V_{\v{q}}$ represents double gate-screened Coulomb interactions with gate distance $d = \SI{25}{nm}$ and relative permittivity $\epsilon_R = 10$, and $A$ is the sample area. The charge density at wavevector $\v{q}$ is given by $\v{\rho}_{\v{q}} = \sum_{\v{k}} \hat{\v{f}}^\dagger_{\v{k}} \braket{u_{\v{k}}|u_{\v{k}+\v{q}}} \hat{\v{f}}_{\v{k}+\v{q}}$ in terms of the periodic part of the Bloch wavefunctions of $h_{\mathrm{TBG}}$. The counterterm $h_{\mathrm{counter}}(\v{k})$ is discussed below.

We partition the Hilbert space into the space of the active band $\mathcal{A}$ and the remote bands $\mathcal{R}$. As the active bands are relatively well-separated, we make the assumption that the ground state density matrix $\hat{\rho}_{\mathrm{full}}$ factorizes into a remote part and an active part:
\begin{equation}
    \hat{\rho}_{\mathrm{full}} = \hat{\rho}_{\mathcal{R}} \otimes \ket{\Psi_{\mathcal{A}}}\bra{\Psi_{\mathcal{A}}}.
\end{equation}
We further assume that $\hat{\rho}_{\mathcal{R}}$ is a product state: fully-filled below the active bands and fully-empty above, which allows us to integrate out the remote bands at mean-field level. Up to a constant this yields
\begin{equation}
    \hat{H}_{\mathcal{A}} = \operatorname{tr}_{\mathcal{R}}
    \Big[\hat{H}_{\mathrm{full}} \hat{\rho}_{\mathcal{R}}\Big]
    = 
    \sum_{\v{k} \in \mathrm{BZ}} \hat{\v{c}}^\dagger_{\v{k}} 
    \Big[
    h_{\mathrm{TBG}}(\v{k}) 
    + h_{\mathrm{HF}}[P_{\mathcal{R}}](\v{k}) 
    - h_{\mathrm{counter}}(\v{k})
    \Big]
    \hat{\v{c}}_{\v{k}}
    + \frac{1}{2A} \sum_{\v{q} \in \mathbb{R}^2} V_{\v{q}} : \hat{\v{\rho}}_{\v{q}}^{\mathcal{A}} \hat{\v{\rho}}_{-\v{q}}^{\mathcal{A}}:,
\end{equation}
where $\hat{\v{c}}$ is the restriction of $\hat{\v{f}}$ to the flat bands, and $\hat{\v{\rho}}_{\v{q}}^{\mathcal{A}}$ is the corresponding flat band density operator. Here $h_{\mathrm{HF}}[P_{\mathcal{R}}]$ is the standard Hartree-Fock Hamiltonian corresponding to the correlation matrix $[\mathcal{P}_R(\v{k})]_{bb'} = \operatorname{Tr}[\rho_{\mathcal{R}} \hat{f}^\dagger_{\v{k} b}\hat{f}_{\v{k} b'}]$, which is the identity matrix on remote bands below the active band, and identically zero otherwise \cite{bultinck2020ground,parker2021field}. (Spin and valley indices are left implicit.) Physically, $h_{\mathrm{HF}}[P_{\mathcal{R}}]$ encodes a background charge density from the filled Fermi sea that affects active electrons. We use $\hat{H}_{\mathcal{A}}$ as the effective Hamiltonian for the active bands.

The final ingredient is the counterterm \cite{bultinck2020ground,Soejima2020_efficient, hofmann2022fermionic, parker2021field}. A counterterm is needed because $h_{\mathrm{HF}}[P_{\mathcal{R}}]$ diverges unphysically with the number of bands. The root of this issue is that some parameters of the BM model already take interactions into account, such as the experimentally-derived Fermi velocity, leading to an unphysical double-counting of some Coulomb interactions. In principle, ``ultraviolet" Hamiltonian $h_{\mathrm{TBG}} - h_{\mathrm{counter}}$ should be fixed by matching to \textit{ab initio} or experimental observations in the ``infrared". (See also \cite{vafek2020renormalization}.) We use the ``decoupled" subtraction scheme: the counterterm is given by the half-filled state of two decoupled layers of graphene. Explicitly, suppose $h^0_K = \operatorname{diag} \begin{pmatrix}
    h_+ & h_-
\end{pmatrix}$ [c.f. Eq. \eqref{eq:single_particle_strained_TBG}] and is diagonalized as $h^0 U^0 = \epsilon^0 U^0$. Then define
\begin{equation}
    h_{\mathrm{counter}} = h_{\mathrm{HF}}[\Delta]; \quad \Delta_{bb'}(\v{k}) = U_0^\dagger \,  P_{\text{CNP,decoupled}} \, U^0
\end{equation}
where $P_{\text{CNP,decoupled}}$ is the diagonal density matrix of graphene at half-filling. The Hartree-Fock correction to the Hamiltonian is therefore
\begin{equation}
    h_{\mathrm{HF}}[P_{\mathcal{R}}](\v{k}) 
    - h_{\mathrm{counter}}(\v{k})
    = h_{\rm{HF}}[P_{\mathcal{R}} - \Delta].
\end{equation}
As $P_{\mathcal{R}}$ and $\Delta$ are approximately equal far from charge neutrality, we make the approximation that bands very far from the Fermi level are irrelevant \cite{bultinck2020ground}. In practice we retain $5$ bands above and $5$ bands below charge neutrality in $P_{\mathcal{R}} - \Delta$ (times valley and spin), and just the two ``flat" bands in $h_{\rm{HF}}[P_{\mathcal{R}} - \Delta]$. 
(Another popular choice is ``infinite temperature subtraction", which leads to $h_{\rm{HF}}[P_{\mathcal{R}} - \Delta] = h_{\rm{HF}}[-I_{\mathcal{A}}/2]$, see e.g \cite{parker2021field}). (Ref. \cite{Kwan2021} checked that the phenomenology of IKS depends only weakly on the subtraction scheme.)

Altogether, our Hamiltonian for the active bands is
\begin{equation}
     \hat{H}_{\mathcal{A}} 
     =
     \sum_{\v{k} \in \mathrm{BZ}} \hat{\v{c}}^\dagger_{\v{k}} 
    h_0'(\v{k})
    \hat{\v{c}}_{\v{k}}
    + \frac{1}{2A} \sum_{\v{q} \in \mathbb{R}^2} V_{\v{q}} : \hat{\v{\rho}}_{\v{q}}^{\mathcal{A}} \hat{\v{\rho}}_{-\v{q}}^{\mathcal{A}}:,\quad
    h_0'(\v{k}) = \Big[
    h_{\mathrm{TBG}}(\v{k}) 
    + h_{\rm{HF}}[P_{\mathcal{R}} - \Delta]
    \Big]
    \label{eq:active_band_many_body_Hamiltonian}
\end{equation}      
As 2-fermion terms may be shuffled between the dispersion and interaction by, e.g., changing the normal ordering reference, $h_0'(\v{k})$ cannot be interpreted directly as a dispersion. However, relative changes in the bandwidth of $h_0'(\v{k})$ are meaningful.

We note that if one defines $\widehat{\delta\rho}_{\v{q}}$ to be the charge density measured relative to (a reference density matrix at) charge neutrality, then the Hamiltonian takes the convenient ``strong-coupling" form $     \hat{H}_{\mathcal{A}} 
     =
     \sum_{\v{k} \in \mathrm{BZ}} \hat{\v{c}}^\dagger_{\v{k}} 
    h_{\mathrm{TBG}}(\v{k}) 
    \hat{\v{c}}_{\v{k}}
    + \frac{1}{2A} \sum_{\v{q} \in \mathbb{R}^2} V_{\v{q}} : \widehat{\delta\rho}_{\v{q}} \widehat{\delta\rho}_{-\v{q}}:$  \cite{bultinck2020ground}.

\subsection{Strain-induced bandwidth and the ``Hartree Peak"}

\begin{figure}
    \centering
    \includegraphics[width=\textwidth]{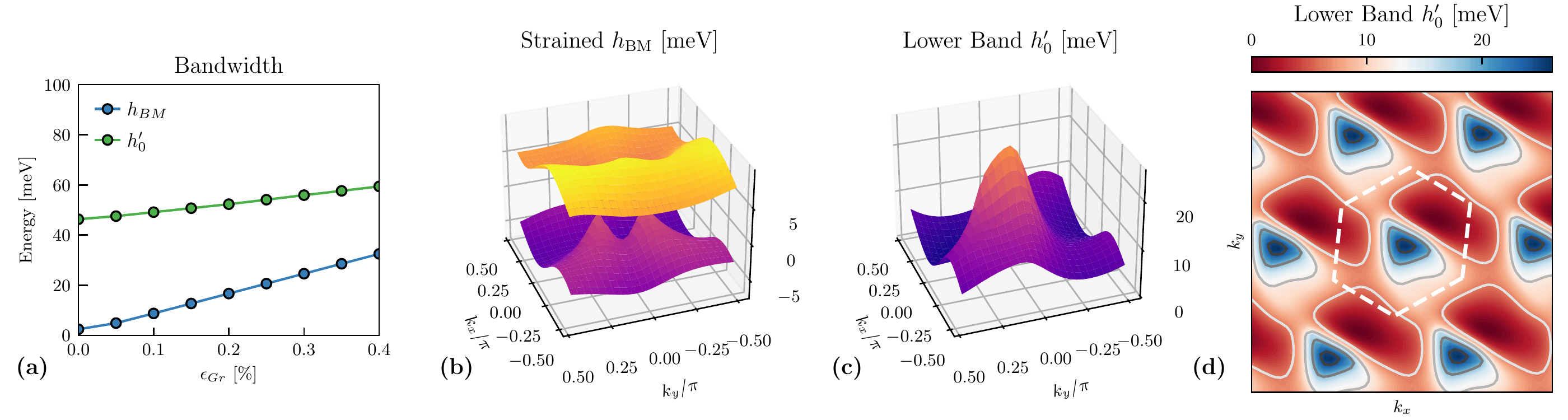}
    \caption{Single-particle energetics of our interacting model. Here $h_{\rm{BM}}$ refers to the strained BM model and $h_0'$, defined in Eq. \eqref{eq:active_band_many_body_Hamiltonian} is related to the total dispersion.
    (a) The bandwidth of $h_{\mathrm{TBG}}$ and $h_{0}'$ as a function of strain. Note the bandwidth increase dramatically with $\eg$. (b) The two flat bands of the BM model for $\eg = 0.2\%$ in meV (c) The lowest band of $h_{0}'$ for $\eg = 0.2\%$ showing the ``Hartree peak" (the upper band is similar). (d) Same as (c), but as a contour plot. This ``cat's eye" pattern will appear below in the normal metal phase. The dashed white hexagon shows the moir\'e Brillouin zone, which is distorted by strain. Parameters match Table \ref{tab:parameters}.
    }
    \label{fig:app_hartree_peak}
\end{figure}

A key effect of heterostrain is to greatly increase the bandwidth. We note that ``bandwidth" is not a strictly well-defined concept in strongly interacting systems, as one can always shuffle 2-fermion terms between the ``dispersion" and ``interaction" parts of the Hamiltonian. Here we consider the ``dispersion`` to be $h_0' = h_{\rm{BM}} + h_{\rm{HF}}[P_{\rm{remote}} - \Delta_{\rm{counterterm}}]$. Fig.\,\ref{fig:app_hartree_peak}(a) shows the bandwidth of the single-particle Hamiltonians $h_{\mathrm{TBG}}$ and $h_{0}'$ as a function of strain. One can see that the bandwidth increases significantly with strain. For comparison, the interaction scale is ${\sim}\,\SI{50}{\milli\electronvolt}$.

Fig.\,\ref{fig:app_hartree_peak}(b) shows $h_{\mathrm{TBG}}$ at $\eg = 0.2\%$. The bandwidth has increased to ${\sim}\,\SI{33}{\milli\electronvolt}$ from  ${\sim}\,\SI{2.5}{\milli\electronvolt}$ at $\eg = 0$. As $C_3$ is broken, the Dirac nodes are no longer pinned to $K^{\pm}$, and instead migrate inwards to the vicinity of $\Gamma$. (Recall that the fragile topology of each valley prevents the Dirac nodes within each valley from being gapped out unless $C_{2}\mathcal{T}$ is broken.) 

Fig.\,\ref{fig:app_hartree_peak}(c,d) show the lower band of $h_{0}'$ at $\eg =0.2\%$. The spectrum is dominated by the ``Hartree peak" of ${\sim}\,\SI{30}{\milli\electronvolt}$ near the $\Gamma$ point. This peak is due to the spatial structure of the wavefunctions \cite{xie2021weak}. The Hartree potential can be thought of as an inhomogeneous background charge density from the filled Fermi sea, which happens to be peaked in the AA region of each unit cell. Holes (electrons) added at generic momentum will be attracted (repelled) from this charge density, thereby lowering (raising) the quasiparticle energy. However, $\psi_{\Gamma}(\v{r})$ vanishes in the AA region at $\eg = 0$ due to $C_3$ symmetry, and is generically small in the AA region even in the presence of strain. Therefore the energies at the $\Gamma$ point are essentially unaffected by the Hartree potential, causing the large Hartree peak (dip) for hole (electron) doping relative to charge neutrality.

Any order that can avoid populating electrons near $\Gamma$ therefore gains a large energetic advantage. Metallic states can always do this by forming a Fermi surface partially up the peak, but we shall see below that IKS and spin spirals can also take advantage of this structure.

\subsection{Computational Basis \& Numerical Details}

We now specify our computational basis. We start from the band of energy bands of $h_{\mathrm{TBG}}$: $\hat{c}_{\v{k} b \tau s}$, where $b \in \{1,2\}$ labels the flat bands. We apply a change of basis
\begin{equation}
    \hat{c}_{\v{k} \sigma \tau s}^\dagger = U_{\sigma b}(\v{k}) \hat{c}_{\v{k} b \tau s}^\dagger,
    \quad 
    \hat{c}_{n, k_y, \sigma \tau s}^\dagger = \int \frac{dk_x}{\sqrt{G_x}} e^{-i\v{k}\cdot \v{a}_1 n} \hat{c}_{\v{k} \sigma \tau s}^\dagger, 
    \label{eq:computational_basis}
\end{equation}
where $U(\v{k})$ is a $2\times2$ unitary that adjusts the gauge so that $\hat{c}_{n, k_y, \sigma \tau s}^\dagger$ create hybrid Wannier orbitals: maximally localized along $\v{a}_1$, but plane-waves with definite momentum $k_y$ \cite{Soejima2020_efficient}. We have taken a rectangular moir\'e Brillouin zone $G_x \times G_y$, and $U(\v{k})$ is complex conjugated for the $K'$ valley so that time-reversal is preserved. At the same time, $U(\v{k})$ transforms to the ``sublattice" basis $\sigma \in \{A,B\}$. It has the property that the microscopic sublattice operator, defined in the band basis as $I_{bb'}(\v{k}) = \braket{u_{\v{k} b}| \sigma^z | u_{\v{k} b'}}$ is diagonal with eigenvalues $\pm \gamma\approx \pm 1$, which is always possible \cite{bultinck2020ground}. We underscore that Eq.\eqref{eq:computational_basis} is a computational basis for all eight active bands, all of which will be dynamical in our model.

Let us reiterate the symmetries of our model. Each valley has a separate $U(1)$ electric charge and $SU(2)$ spin conservation, which gives a combined $U(2) \times U(2)$ continuous symmetry (at the level of the Lie algebra). We also have time-reversal $\mathcal{T}$, $C_{2z}$ and the anti-unitary combination $C_{2z}\mathcal{T}$, which is $k$-local. Of course, we also have moir\'e scale translation symmetry $T_{\v{a}_i}$ along $\v{a}_{1,2}$. Maximal localization along $\v{a}_1$ implies the Wannier orbitals are eigenstates of the projected position operator $\mathcal{P} (\v{a}_1 \cdot \hat{\v{r}}) \mathcal{P}$, where $\mathcal{P}$ projects to the active bands. Finally, we note that the BM Hamiltonian is approximately particle-hole symmetric, but strain breaks particle-hole symmetry further. The full symmetry content of the model is discussed in Ref. \cite{bultinck2020ground}.

\begin{table}[]
    \begin{center}
    \begin{ruledtabular}
    \begin{tabular}{ll}
        Parameter & Value(s) \\[0.2em] \colrule
        Twist angle $\theta$ & $\ang{1.08}$ \\
        Strain $\eg$ & $0\%$ -- $0.2\%$\\
        Chiral parameter $\kappa$ & 0.5 -- 0.8 \\
        Interlayer tunnelling $w_{AB}$ &  \SI{110}{\milli\electronvolt} \\
        Gate distance  $d$ & \SI{25}{\nano\meter} \\
        Relative permitivity $\epsilon_r$ & 10 \\[0.3em]
        \colrule
        Active bands (per valley per spin) & 2\\
        Subtraction Scheme (see text) & Decoupled\\
        Remote bands for subtraction (per valley per spin) & 10\\
        Cylinder circumference $L_y$ & $4,6$ \\
        MPO accuracy $\epsilon_{\mathrm{MPO}}$ & $10^{-2}$ \si{\milli\electronvolt} \\
        MPO bond dimension $\chi_{\mathrm{MPO}}$ & $250 - 1500$ \\
        State bond dimension $\chi$ & $1024 - 24,576$ \\[0.3em]
    \end{tabular}
    \end{ruledtabular}
    \end{center}
    \caption{Parameters our model and DMRG calculation.}
    \label{tab:parameters}
\end{table}

\section{Details of the DMRG Calcuations}
\label{app:DMRG_details}
This section gives numerical details of our DMRG computations, including flavor polarization, and defines the correlation functions we study in the main text and below.

In practice we select $L_y$ cuts through the Brillouin zone at evenly spaced momenta $k_y = \frac{2\pi m}{L_y}$ for integer $m$. We then resolve $\hat{H}_{\mathcal{A}}$ in the computational basis as an matrix product operator (MPO). Naively, such an MPO would have bond dimension $\chi_{\mathrm{MPO}} \approx 100,000 - 300,000$ (typical MPOs are $\chi_{\mathrm{MPO}} \approx 60$ for short-range 2D systems). We apply MPO compression \cite{parker2020local,Soejima2020_efficient} to reduce the MPO bond dimension to $\chi_{\mathrm{MPO}} < 1500$, while retaining an accuracy of $\epsilon_{\mathrm{MPO}} =$\SI{10e-2}{meV} or better, making the computation tractable. We note that DMRG simulations have memory requirements $O(\chi_{\mathrm{MPO}} \chi^2)$ where $\chi$ is the bond dimension of the state (hundreds or thousands of gigabytes of memory for our largest simulations).

We use two ``rings" around the cylinder as a unit cell for infinite DMRG. This allows breaking of translation $T_{\v{a}_1}$ along the cylinder down to $2T_{\v{a}_1}$ (possibly times a phase), but enforcing $k_y$ momentum conversation prevents translation breaking around the cylinder. Our ansatz always permits $T_{\v{a}_1}$ to be modified by a phase factor.

\subsection{DMRG Flavor Sectors}

Our simulations explicitly conserve a $U(1)^4 \subset U(2) \times U(2)$ symmetry, corresponding to:  total electric charge, valley charge, and spin-$z$ in each valley, respectively. These are measured by $\hat{n}$, $\tau^z$, $s^z_{K}$, and $s^z_{K'}$ respectively. As usual, the Hamiltonian is block-diagonal, with sectors labelled by the integer charges of these four symmetries. DMRG will find a ground state within each sector, based on the charge of the initial state. We consider the four ``least charged" sectors, which should contain the global ground state. These are labeled by their distinct charge sectors $(\tau^z,s^z)$. Here $s^z = s^z_K + s^z_{K'}$ is the sum of the spins in both valleys. We only consider the diagonal sectors with $s^z_{K} = s^z_{K'}$. We work at filling $\nu=-3$, which fully specifies the electric charge sector. Explicitly, we consider the following four sectors:
\begin{enumerate}
    \item $(\tau^z,s^z) = (1,1)$: electrons are populated only in the $\tau=K, s=\uparrow$ sector.
    \item $(\tau^z,s^z) = (0,1)$: electrons are equally populated between the two $\tau=K/K', s=\uparrow$ sectors.
    \item $(\tau^z,s^z) = (1,0)$: electrons are equally populated between the two $\tau=K, s=\uparrow/\downarrow$ sectors.
    \item $(\tau^z,s^z) = (0,0)$: electrons are equally populated between all four $\tau=K/K', s=\uparrow/\downarrow$ sectors.
\end{enumerate}

\subsection{2D Correlators}

The key tool to identify phases in DMRG is correlation functions. In simple cases, phases have well-defined order parameters that are two-point correlation functions. However, even simple phases in 2D can be difficult to diagnose on a quasi-1D cylinder due to enforced algebraic order from the  Hohenburg-Mermin-Wagner theorem. These must be characterized by the \textit{trend} of correlators as a function of the bond dimension. This section introduces the 2D correlators whose 1D versions are evaluated in DMRG. 

A key correlator is the 2-electron correlation matrix
\begin{equation}
    P_{\beta\alpha}(\k)=  \braket{ c^\dagger_{\k,\alpha} c_{\k, \beta}},
    \label{eq:P_correlation_matrix_defn}
\end{equation}
an $8\times 8$ matrix for each $k$, where $\alpha,\beta \in \st{\sigma,\tau,s}$ index sublattice, valley, and spin as above. We frequently consider the diagonal component of this correlator, such as $n_{\v{K}\uparrow}(\v{k}) = \sum_{\sigma} \braket{ c^\dagger_{\k,\sigma,K,\uparrow} c_{\k, \sigma,K,\uparrow}}$.

To detect IKS order and other phases, it is frequently useful to consider $k$-nondiagonal operators. Define 
\begin{equation}
    \hat{\Delta}_{\q}^{ijk}(\v{k}) = \hat{\v{c}}_{\k+\q}^\dagger \sigma^i \tau^j s^k \hat{\v{c}}_{\k},
\end{equation}
which picks out a particular combination of Pauli matrices in flavor space. To convert from 2D to the quasi-1D cylinder, we apply a Fourier transform $\hat{c}_{\k, \sigma\tau s}^{\dagger}:=\sum_{n \in \mathbb{Z}} e^{i \left(\k\cdot n\v{a}_1\right)} \hat{c}_{n, k_{y}, \sigma\tau s}^{\dagger}$ where $n \in \mathbb{Z}$ labels the hybrid Wannier orbital centered at $n \v{a}_1$ along the cylinder. Then 
\begin{equation}
    \hat{\Delta}_{\q}(\v{k}) = \sum_{m,n \in \mathbb{Z}} e^{i \v{k} \cdot m \v{a}_1} \hat{\Delta}_{n,q_y}(m,k_y);
    \quad \hat{\Delta}^{ijk}_{n,q_y}(m,k_y) = \hat{c}^\dagger_{n+m, k_{y}+q_{y}}\sigma^i\tau^j s^k \hat{c}_{n, k_{y}}.
    \label{eq:2_point_operator_in_wannier_basis}
\end{equation}
Therefore, if some 2D order operator $\braket{\hat{\Delta}_\q (\k)}$ is peaked at $\q=\q_0$, then we expect $\braket{\hat{\Delta}_{n,q_{y}} (m, k_y)}\sim \delta(q_{0,y})e^{-in(\q_0\cdot\v{a}_1)}$, and the two components of $\q_0$ can be extracted accordingly. When taking the quasi-1D limit, the Hohenberg-Mermin-Wagner theorem ensures that $U(1)$ symmetries (e.g.) are not spontaneously broken. So, if $\hat{\Delta}$ an order parameter for a continuous symmetry $G$ in 2D, in the quasi-1D (thin cylinder) limit we expect either algebraic or exponential decay (depending on $G$) of $\braket{\hat{\Delta}_n \hat{\Delta}_0}$ along the cylinder. This becomes a 4-point correlation function
\begin{equation}
    C_{\Delta}(n,q_y) = L_y^{-2}\Braket{\sum_{k_y}\Delta_{0,q_{y}} (0, k_y) \sum_{k'_y}\Delta^\dagger_{n,q_{y}} (0, k'_y)}.
\end{equation}
When symmetry breaking is present, we expect $C_{\Delta}(n,q_y)\sim\delta(q_{0,y})e^{in(\q_0\cdot\v{a}_1)}$, allowing us to extract the 2D $\v{q}$ from DMRG data.

At finite bond dimension $\chi$, the correlator must decay exponentially at the largest scales: $C_\Delta(n,q_y) \sim e^{-n/\xi_{\Delta}}$. If the correlator is algebraic, then we expect a divergence $\xi_{\Delta} \to \infty$ as $\chi \to \infty$. In fact, the subleading eigenvalues of the transfer matrix will diverge as well, which may be used to assess scaling relations in extremely high-bond dimension data \cite{vanhecke2019scaling}. For spin-polarized IKS order, which breaks $U(1)_{\rm{valley}}$ we indeed expect such algebraic decay $\n{C_\mathrm{IKS}(n,q_y)}\sim n^{-\eta(L_y)}$ as $\chi \to \infty$. As $L_y \to \infty$, $\eta(L_y) \to 0$, recovering the 2D limit. In complex systems such as the one under consideration here, one cannot reliably determine if a given correlation length $\xi_{\Delta}$ is truly diverging or simply approaching a finite value slowly as a function of $\chi$. Nevertheless, we qualitatively observe that there often is a ``scaling regime": a sufficiently large $\chi$ after which the behavior $\xi_{\Delta}(\chi)$ either plateaus or grows regularly. We take such behavior as a indication that our bond dimension is sufficiently large to capture the ``true" ground state.

\section{Phase diagram analysis at flavor-polarized sectors}
\label{app:phase_identification}
In the following two sections, we will discuss the various phases found within DMRG and indicate the resulting phase diagram. This section focus on the phases \textit{not} in the $(\tau^z,s^z) = (0,0)$ sectors, where certain choices of flavor polarization are enforced.

\subsection{Quantized Anomalous Hall order}
\label{app:QAH}

This section will discuss the quantized anomalous Hall (QAH) order that appears at low strain in all sectors. The QAH insulator is a ground state within strong coupling theory in the chiral flat limit at $\n{\nu} = \pm 1, \pm 3$ \cite{bultinck2020ground, lian2020tbg}. The simplest example is described by the correlation matrix $P_{\mathrm{QAH}} = (\frac{1}{2}[1+\sigma^z])(\frac{1}{2}[1+\tau^z])(\frac{1}{2}[1+s^z])$, which entirely fills the $(A,K,\uparrow)$ sector. This state has full polarization of the QAH order parameter
\begin{equation}
C_{\mathrm{QAH}} = \frac{1}{N_k} \sum_\k \braket{\hat{\v{c}}^\dagger_{\k}\sigma^z\tau^z \hat{\v{c}}_{\k}}.\\
\label{eq:QAH_order_parameter}
\end{equation}
As $C=\sigma^z\tau^z$ measures the Chern number, states with $C_\mathrm{QAH} = \pm 1$ correspond to full polarization in Chern bands, and are thus Chern insulators with $C=\pm 1$. Even when $C_\mathrm{QAH}$ is away from polarization, we expect the state is adiabatically connected to the fully polarized Chern insulators. We therefore use  $\mathrm{sign}(C_\mathrm{QAH})$ as a signature of the Hall conductance of the state. We also define
\begin{equation}
    C_{\sigma^+} = \frac{1}{N_k} \sum_\k \braket{\hat{\v{c}}^\dagger_{\k}\sigma^+ \hat{\v{c}}_{\k}},
    \label{eq:C_sigma_plus}
\end{equation}
which is small in the QAH phase, but will appear in the semimetal discussed below.

The chiral flat limit features an emergent $U(4)_+ \times U(4)_-$ symmetry acting within each Chern sector \cite{bultinck2020ground}. Acting with any $U_+ \in U(4)_+$ on $P_{QAH}$ gives another ground state in this limit. For instance, the state $P_{\mathrm{QAH-IVC}} = \frac{1}{2}\left(I + \sigma^x \tau^x\right) (\frac{1}{2}[1+s^z])$ with equal occupations in $(A,K,\uparrow)$ and $(B,K',\uparrow)$. Working perturbatively in deviations from the chiral limit, one expects that $P_{\rm{QAH}}$ and $P_{\rm{QAH-IVC}}$ are still competitive ground state candidates at $\kappa\approx 0.65$, and should still be nearly degenerate. This is borne out in our DMRG results.

\begin{figure}[t]
    \centering
    \includegraphics[width=0.55\textwidth]{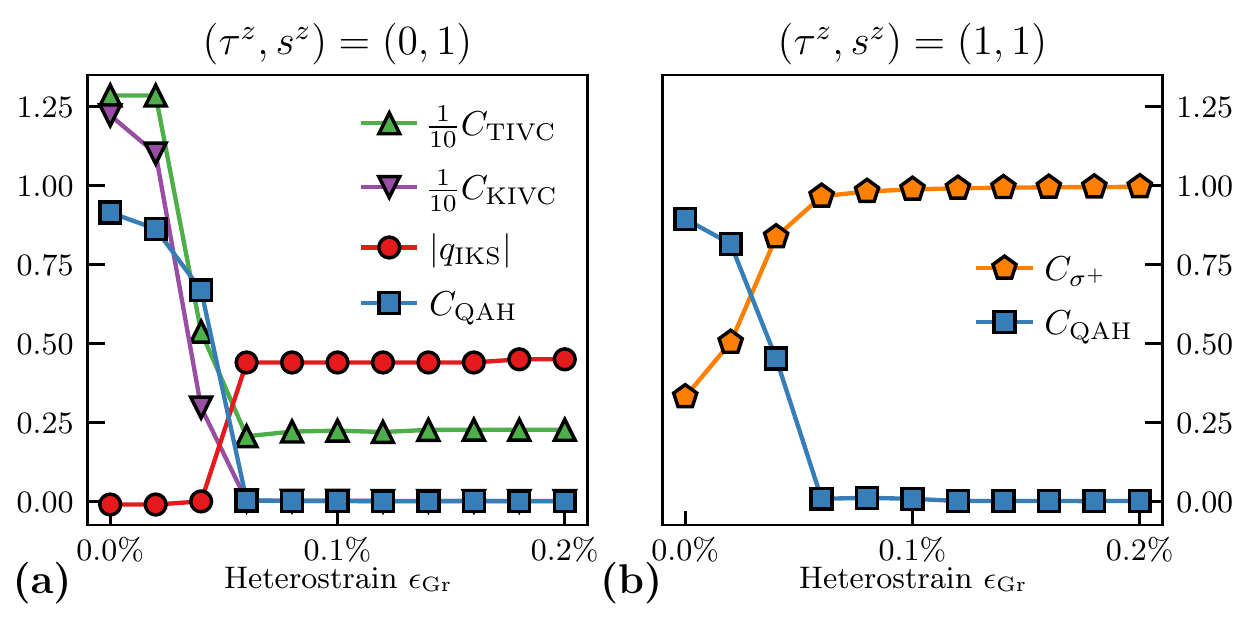}
    \caption{(a) Correlation functions and order parameters in the $(0,1)$ sector as a function of $\eg$. See text for details. (b) Correlation functions in the $(1,1)$ sector as a function of $\eg$. We note that the phase boundaries change slightly between the $L_y = 6$ data here and $L_y = 4$ results shown in the main text. Parameters: $\chi=3072$, $\kappa=0.7$, $L_y=6$, otherwise matching Table \ref{tab:parameters}.
    }
    \label{fig:app_QAH}
\end{figure}

We find that the ground state of each sector is a variant on QAH, adapted for that particular symmetry sector. For simplicity, we focus on the $(0,1)$ and $(1,1)$ sectors where spin does not play a role; the other sections are similar. Fig.\,\ref{fig:app_QAH} shows the value of the order parameter $C_{\mathrm{QAH}}$ as a function of $\eg$. It is nearly unity at $\eg =0$ for both sectors, but strain quickly drives a transition to $C_{\mathrm{QAH}} = 0$. For the $(1,1)$ sector, the state we find is adiabatically connected to $P_{\rm{QAH}}$, similar to previous reports \cite{Soejima2020_efficient}. For $(\tau^z, s^z) = (0,1)$, however, the requirement that occupations are balanced between valleys excludes the $P_{\rm{QAH}}$ state; the $(0,1)$ state is instead descended from $P_{\rm{QAH-IVC}}$. In 2D, this breaks $U(1)_{\rm{valley}}$ symmetry. In 1D, however,  $U(1)$ symmetry cannot be spontaneously broken; instead algebraic order will manifest in intervalley correlation functions. Define operators
\begin{align}
\hat{\Delta}^{\rm{TIVC}}(m,k_y) = \hat{c}^\dagger_{m,k_y} \sigma^x \tau^x \hat{c}_{0,k_y}, \quad 
\hat{\Delta}^{\rm{KIVC}}(m,k_y) = \hat{c}^\dagger_{m,k_y} \sigma^x \tau^y \hat{c}_{0,k_y},
\label{eq:TIVC_KIVC_operators}
\end{align}
and the corresponding correlators $C_{\rm{TIVC/KIVC}} := \sum_{n \in \mathbb{Z}} |\braket{\hat{\Delta}(n,0) \hat{\Delta}(0,0)}|^2$ \cite{Parker2021}. Fig.\,\ref{fig:app_QAH} (a) shows both $C_{\rm{TIVC/KIVC}}$ are large in the low-$\eg$ phase. In fact, the intervalley correlation length diverges with $\chi$, indicating the expected algebraic correlations. We note that TIVC and KIVC correlations are not distinct when only a single sublattice is occupied. However, note that after the transition, $C_{\rm{KIVC}}$ approaches zero while $C_{\rm{TIVC}}$ remains finite, a fact that will return below in our discussion of IKS order.

We therefore identify the order in the $(1,1)$ sector as QAH order, and the $(0,1)$ sector as QAH with intervalley coherence (QAH-IVC). The $(1,0)$ sector is consistent with the ferromagnet in $xy$-plane, similar to the 1D Heisenberg chain discussed in App.\,\ref{app:heisenburg_analogy}. We treat $(0,0)$ sector, which shows spinful QAH-IVC order, separately in App.\,\ref{app:spinfulIKS}. In fact, Fig.\,\ref{fig:phase_diagram} of the main text shows all four sectors have similar energies.

\subsection{Nematic Semimetallic Order}
\label{app:CNSM}

This phase describes the large-strain phase in the $(1,1)$ sector, which we identify as a nematic semimetal, closely resembling the nematic semimetal discovered previous in the strong coupling regime \cite{Soejima2020_efficient, Kwan2021, liu2021_NSM}. Previous studies have shown the nematic semimetal phase to be a competitive ground state candidate favored by $\mathcal{C}_3$ symmetry-breaking terms, in our case the uniaxial heterostrain. The nematic semimetal descends from a ``parent state" within strong-coupling theory after the application of a singular gauge transformation which eliminates the Berry flux \cite{khalaf2020_softmodes}. At $\eg = 0$, the gap to the nematic soft mode decreases with $\kappa$, eventually driving a transition from QAH to the nematic semimetal \cite{Soejima2020_efficient,khalaf2020_softmodes}. Adding a small amount of strain, which breaks $C_3$ explicitly, has a similar effect \cite{Kwan2021, Parker2021}.

We examine the state at $\eg = 0.15\%$, $\kappa =0.7$ in detail. As we are in the $(1,1)$ sector, electrons only occupy the $K$-valley with spin $\uparrow$. The first observation is the state is quite close to a Slater determinant. To assess this, we define the single-particle Shannon entropy
\begin{equation}
    S_{\mathrm{Sh}} = - \frac{1}{N_k} \sum_{k} \Tr[P(k) \ln P(k)],
    \label{eq:Shannon_entropy}
\end{equation}
where $P(k)$ is the 2-body correlation matrix defined in Eq. \eqref{eq:P_correlation_matrix_defn}. A state may be represented as a Slater determinant if and only if $P(k)$ is a projector i.e. $P(k)^2 = P(k)$. In this case, its eigenvalues are either $0$ or $1$, so $\Tr[P(k) \ln P(k)] =0$ for all $k$. One may therefore use $S_{\mathrm{Sh}}$ to assess how far a given state is from a Slater determinant. In this case, $S_{\mathrm{Sh}} \approx 0.02$, so the state is extremely close to a Slater determinant, and we may understand the state through single-particle considerations.

To better display the physics of this phase, we employ a ``superresolution" technique. Namely, we thread flux $\phi$ through the cylinder so that momentum cuts move to $k_y(m,\phi) = \frac{2\pi (m+\phi)}{L_y}$ for $0 \le \phi < 1$. So long as no transitions occur as a function of $\phi$ (which holds here, but is non-generic), we may combine data from DMRG runs at multiple $\phi$ to improve our $k_y$ resolution significantly. 

\begin{figure}[t]
    \centering
    \includegraphics[width=1\linewidth]{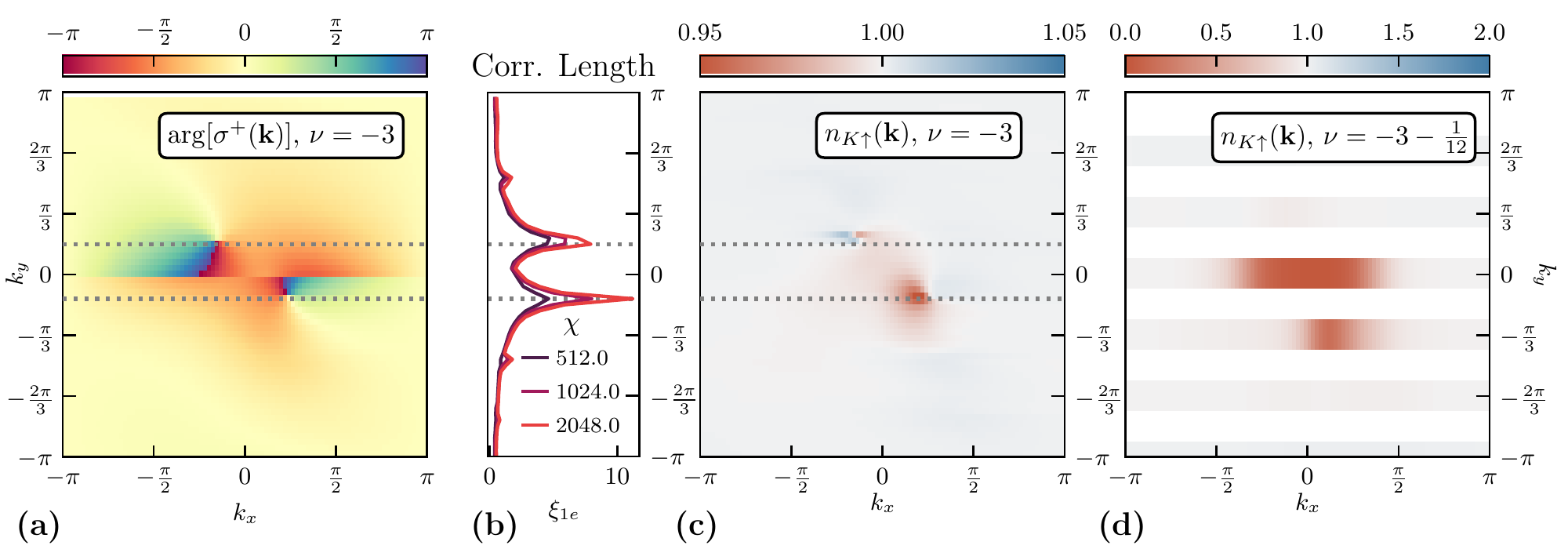}
    \caption{Details of the CNSM phase. Panels (a,b,c) combine multiple DMRG runs with different flux threadings to produce ``superresolution" plots. (a) Shows the phase of the Chern off-diagonal correlations over the Brillouin zone. The phase winds by $2\pi$ around each Dirac node. The fragile topology ensures the gauge cannot be both smooth and periodic; we adopt a periodic gauge with a discontinuity at $k_y =0$. Dashed grey lines indicate the $k_y$ of the Dirac nodes. (b) Single-electron correlation lengths as a function of $k_y$ for various $\chi$'s, which diverge at the Dirac nodes. (c) Electron occupations in the $K\uparrow$ sector. Note small electron/hole pockets around the Dirac nodes. (d) Electron occupations in the $K\uparrow$ sector at filling $\nu=-3 -\frac{1}{12}$. The hole pocket has expanded greatly. Parameters: $\chi=2048$, $\eg = 0.15\%$, $\kappa =0.7, L_y = 6$, $(\tau^z,s^z) = (1,1)$ sector.}
    \label{fig:semimetal}
\end{figure}

Fig.\,\ref{fig:semimetal}(c) shows the electron occupations across the Brillouin zone. Since the state is Slater-like, we can interpret this in terms of an effective Hartree-Fock bandstructure of two bands. The electron density is almost uniform, which indicates the state fills one of the two bands. However, there are additional features near $\Gamma$, which are due to Dirac nodes. To see this, Fig.\,\ref{fig:semimetal}(a) shows the phase winding of 
\begin{equation}
\arg[\sigma^+(\v{k})] = \arg{\braket{ \hat{c}^{\dagger}_{\k, A,K, \uparrow} \hat{c}_{\k, B,K,\uparrow}}}.
\end{equation}
For states close to a Slater determinant, this winds by $\pm 2\pi$ around Dirac nodes \cite{Soejima2020_efficient}. Two vortices are clearly visible near $\Gamma$, each with a $+2\pi$ winding. We note that, due to fragile topology, one cannot choose a gauge which is both smooth and periodic. We choose a periodic gauge with a discontinuity across $k_y=0$. Fig.\,\ref{fig:semimetal}(b) shows the electron-electron correlation length $\xi_{1e}$, which diverges at precisely the same $k_y$ values as the vortex centers, indicating gapless electrons. We may conclude that we an effective bandstructure with two Dirac nodes of the same chirality in the vicinity of $\Gamma$ --- quite similar to the phenomenology of the nematic semimetal, but also similar to the underlying bandstructre of TBG in the presence of strain.

In the presence of $C_3$ breaking, the Dirac nodes are not pinned to half-filling of the two bands, and may shift in energy. The slight deviation from uniform filling in Fig.~\ref{fig:semimetal}(c) may thus be a consequence of such effect, though the variation is too small to be conclusive. We note that the occupied band shows Hartree peaks feature, as seen in Fig.~\ref{fig:semimetal}(d): upon hole-doping, the holes enter almost entirely at $\Gamma$. In summary, this state is a semimetal with two Dirac nodes of the same chirality.

\subsection{Commensurate Spin Spiral Order}
\label{app:CSS}

The ground state in the $(1,0)$ sector above a low strain $\eg=0.05\%$ has commensurate spin spiral order. Similar to IKS, this order breaks both $U(1)_{\rm{spin}}$ and moir\'e translation symmetry $\hat{T}_{\v{a}_i}$, but preserves a combined symmetry
\begin{equation}
    \hat{T}^{\mathrm{CSS}}_{\v{a}_i} = \hat{T}_{\v{a}_i} e^{i \q_\mathrm{CSS} \cdot \v{a}_i s^z/2}
\end{equation}
for some offset vector $\qCSS$. We will see that $\qCSS = \v{g}/2$ for some reciprocal lattice vector $\v{g}$. 

\begin{figure}[t]
    \centering    
    \includegraphics[width=\textwidth]{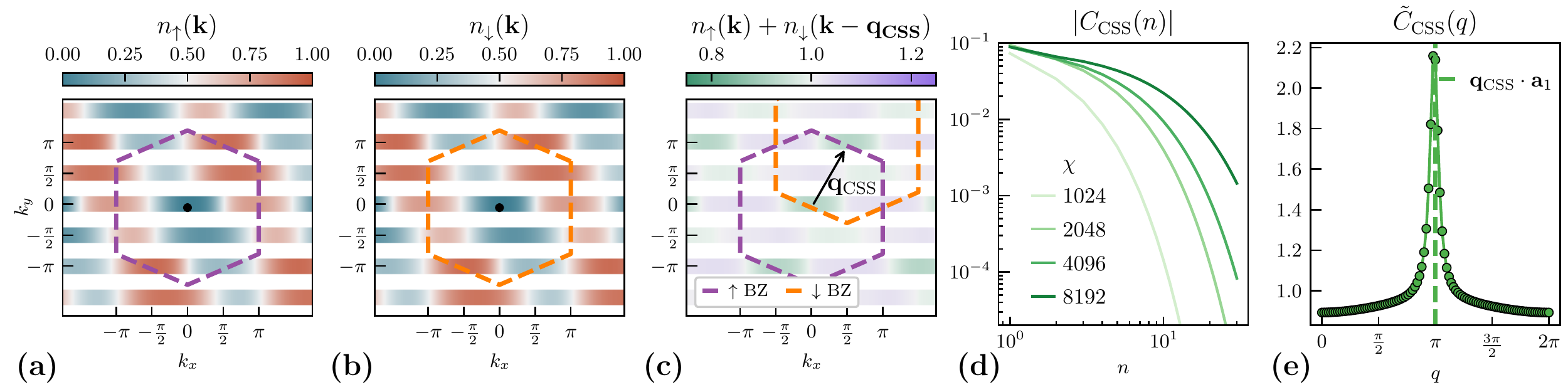}
    \caption{Details of the commensurate spin spiral ordering. (a) Electron occupations for spin $\uparrow$. (b) Electron occupations for spin $\downarrow$. Note these are virtually identical to the previous panel. (c) Offset occupations $n_{\uparrow}(\v{k}) + n_{\downarrow}(\v{k}+\qCSS) \approx 1$. (d) The $CSS$ correlation function, whose correlation length is growing quickly with $\chi$. (e) The discrete Fourier transform of the same correlation function with respect to $n$, which is strongly peaked at $\qCSS \cdot \v{a}_1 = \pi$. Parameters: $\chi=8192$, $\kappa =0.65, \eg =0.2\%, L_y = 4$, $(\tau^z,s^z) = (1,0)$ sector.}
    \label{fig:CSS}
\end{figure}

Fig.\,\ref{fig:CSS} shows the properties of the commensurate spin spiral state [compare to Fig.\,\ref{fig:IKS} in the main text]. Panels (a,b) show the electron occupations for spin $\uparrow,\downarrow$ respectively. Note that the occupations are virtually identical; the state is symmetric under $s_x$-symmetry. Just like in IKS, we observe that
\begin{equation}
    N^{(S)}_{\v{q}_\text{CSS}}(\v{k}) = n_\uparrow(\v{k}) + n_\downarrow(\v{k} - \v{q}_\text{CSS}) \approx 1
    \label{eq:CSS_invariant_occupations},
\end{equation}
where, in this case, $\qCSS = \v{g}_1/2$ is half of a reciprocal lattice vector.

Such commensurate spin spiral ordering will break $U(1)_{\rm{spin}}$, and therefore manifest as an increasing correlation length in the appropriate correlation function. Define the operator
\begin{equation}
        \hat{\Delta}_{\mathrm{CSS}}(\k) = \hat{\v{c}}^\dagger_{\k+\q_\mathrm{CSS}}s^+\hat{\v{c}}_{\k} = \sum_{m, n \in \mathbb{Z}}  e^{i\v{k}\cdot m \v{a}_1} \hat{\Delta}_{n,q_y}(m,k_y); \quad
        \hat{\Delta}^{\rm{CSS}}_{n,q_y}(m,k_y) = \hat{c}^\dagger_{n+m,k_y+q_{\rm{CSS},y}} s^+\hat{c}_{n,k_y},
        \label{eq:CSS_operator}
\end{equation}
and put $C_{\rm{CSS}}(n,q_y) := L_y^{-2} \sum_{k_y,k_y'}\braket{\hat{\Delta}_{0,q_y}(0,k_y) \hat{\Delta}_{n,q_y}(0,k_y')}$. Fig.\,\ref{fig:CSS} (d) shows the absolute value of $C_{\rm{CSS}}(n,q_y = [\v{g}_1/2]_y)$, whose correlation length is increasing quickly with $\chi$, a sign of order that breaks $U(1)_{\rm{spin}}$ but restores the product $\hat{T}^{\mathrm{CSS}}_{\v{a}_i}$. Fig.\,\ref{fig:CSS} (e) shows the discrete Fourier transform of $C_{\rm{CSS}}$ with respect to $n$ along the cylinder, which is strongly peaked at $q = \pi$. Therefore $C_{\rm{CSS}}$ independently finds $\qCSS = \v{g}_1/2$.

Spin-flip symmetry enforces a commensurate offset vector. Given $n_{\uparrow}(\k)=n_{\downarrow}(\k)$, the same state can also be described by CSS order with $-\v{q}_\text{CSS}$. Since $\qCSS$ and $\qCSS + \v{g}$ are equivalent for any reciprocal lattice vector $\v{g}$, we must have $\qCSS = \v{g}/2$ to ensure consistency. This is a key difference between IKS order and CSS order.

Let us note a few other properties of this state. The CSS order described here is \textit{not} spin density wave order (though that will appear in the $(0,0)$ sector considered below). We have explicitly checked that the spin occupations are equal in each ring of the cylinder, as expected for a state with $\hat{T}^{\mathrm{CSS}}_{\v{a}_i}$ symmetry but not for an SDW. As only the $K$ valley is populated, time-reversal symmetry is explicitly broken. However, there is equal occupation in both sublattices, suggesting no net Hall conductance. As noted in the main text, the phenomenology of CSS and IKS order are similar in that both depopulate the region near $\Gamma$ in all sectors. We conclude that the high-strain phase of the $(1,0)$ sector exhibits commensurate spin spiral ordering.

\subsection{Spin-Polarized IKS order}
\label{app:IKS}

This section focuses on the spin-polarized IKS order in the $(\tau^z,s^z) = (0,1)$ sector, expanding the discussion in the main text. As mentioned there, IKS features intervalley correlations with a momentum offset $\q = \qIKS$, which is measured by the operator
\begin{equation}
        \hat{\Delta}_{\mathrm{IKS}}(\k) = 
            \hat{\v{c}}^\dagger_{\k+\q_\mathrm{IKS}} 
                    \sigma^x \tau^+ 
            \hat{\v{c}}_{\k} 
            = 
            \sum_{m, n \in \mathbb{Z}}  e^{i\v{k}\cdot m \v{a}_1} \hat{\Delta}^{\rm{IKS}}_{n,q_y}(m,k_y); \quad
        \hat{\Delta}^{\rm{IKS}}_{n,q_y}(m,k_y) = 
        \hat{c}^\dagger_{n+m,k_y+q_{\rm{IKS},y}} 
            \sigma^x \tau^+
        \hat{c}_{n,k_y}.
\end{equation}
We define the $q_y$-dependent correlation function 
\begin{equation}
    C_{\rm{IKS}}(n,q_y) := L_y^{-2} \sum_{k_y,k_y'}\braket{\hat{\Delta}^{\rm{IKS}}_{0,q_y}(0,k_y) \hat{\Delta}^{\rm{IKS}}_{n,q_y}(0,k_y')} \xrightarrow{n \gg 1} e^{-n/\xi_{\rm{IKS}}[q_y]},
    \label{eq:app_IKS_correlator}
\end{equation}
where $\xi_{\rm{IKS}}[q_y]$ is the IKS correlation length regarding the $q_y$ sector. Since $s^z = 1$, we ignore spin in this discussion; the spinful variant of IKS is discussed below.

\begin{figure}[t]
    \centering
    \includegraphics[width=0.8\linewidth]{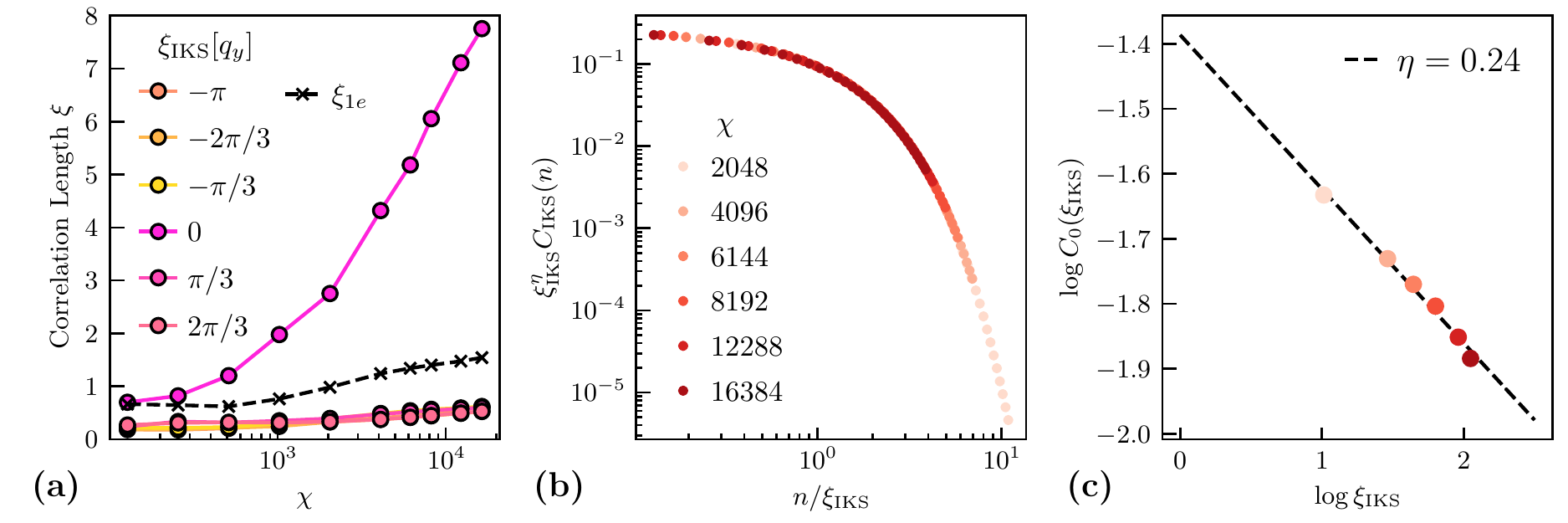}
    \caption{(a) IKS correlation lengths as a function of $\chi$ for each momentum offset $q_y$. The single-electron correlation length $\xi_{1e}$ is shown on the same scale. (b) Scaling collapse of $\CIKS$ via Eq.\,\eqref{eq:IKS_scaling_collapse}. (c) Inference of the scaling exponent $\eta = 0.24$ as described in the text. Parameters: $\chi=16,384$, $\kappa =0.65, \eg =0.2\%, L_y = 6$, $(\tau^z,s^z) = (0,1)$ sector.}
    \label{fig:app_IKS}
\end{figure}

We measure the four-point correlation function $C_{\rm{IKS}}(n,q_y)$ for each $q_y$. Fig.\,\ref{fig:IKS}(d) of the main text shows the $q_y =0$ case, and we plot the evolution of the correlation lengths $\xi_{\rm{IKS}}[q_y]$ with $\chi$ in Fig.\,\ref{fig:app_IKS}(a). For $\chi > 1024$, $\xi_{\rm{IKS}}[q_y = 0]$ is clearly dominant, and is diverging with system size. This matches the expectation that $U(1)_{\rm{valley}}$ breaking in 2D manifests as algebraic order in $C_{\rm{IKS}}(n,q_y)$ on the cylinder, discussed in the main text.

We note that the correlation length can also be extracted from eigenvalues of the MPS transfer matrix. Let $\lambda_{\rm{IKS},q_y}$ be the largest eigenvalues of the MPS transfer matrix in the IKS sector $\Delta Q_\mathrm{IKS} = (\Delta q_\text{electron} = 0, \Delta q_\text{valley} = 2, \Delta k_y = q_y)$. For each $q_y$, these obey $\lambda = e^{-1/\xi}$. Furthermore, $\lambda_{\rm{IKS},q_y=0}$ is the largest eigenvalue in all sectors of the transfer matrix --- and indeed the only substantial one.

We note that single-electron correlation length $\xi_{1e}$, shown in Fig.\,\ref{fig:app_IKS} (a), is also growing as a function of $\chi$. In some cases this behavior is associated with metallic order. However, $\xi_{1e}$ may just be slowly converging to a relatively large value. If the state was metallic, one would likely expect $\xi_{1e}$ (from $n(k) \sim [k-k_F]^\alpha$) or $\xi_{0e}$ (gapless particle-hole modes) to be the dominant correlation length in the system, which is the case in the normal metal below. This data is therefore not sufficient to determine if the state is metallic or insulating but, given that the BZ occupation is uniform after the shift by $\qIKS$, the latter seems more plausible.

Finally, we demonstrate the unified scaling collapse of the IKS correlation functions as
\begin{equation}
    \n{\CIKS(n,\xiIKS)} = \xiIKS^{-\eta} \n{\CIKS(n/\xiIKS,1)},
    \label{eq:IKS_scaling_collapse}
\end{equation}
which is performed according to Ref.\,\cite{Parker2021}. Given that all correlations functions must decays exponentially at finite bond dimensions, we perform fits on $\CIKS(n\gg 1,\xiIKS) \equiv C_0(\xiIKS)e^{-n/\xiIKS}$ to extract the $n$-independent prefactor. If the scaling collapse holds true, we will expect $C_0(\xiIKS)\sim \xiIKS^{-\eta}$, from which the exponent $\eta$ is extracted. Fig.\,\ref{fig:app_IKS}(b-c) shows the fitting results for all correlation functions among $\chi=2048-16384$, where an excellent scalling collapse is established at $\eta=0.24$. This relation predicts an algebraic behavior for $\CIKS(n)$ at the 2D limit: $\CIKS(n,\xiIKS\rightarrow\infty)\sim n^{-\eta}$, which firmly supports that IKS order is robust in our DMRG ground state.

\section{1D Heisenberg model in $S^z$ = 0 sector}
\label{app:heisenburg_analogy}
This section briefly reviews some physics of the Heisenberg model, which will be crucial to understand the spin correlations of TBG in App.~\ref{app:spinfulIKS}. Namely, we will argue there that the following phenomena in the $S^z=0$ sector are indicative of ferromagnetic order:

\begin{enumerate}
    \item Because the state has in-plane long range order, the correlation length for the sector that contains $S^+(x) S^-(0)$ diverges with $\chi$. 
    \item Because the state breaks rotational symmetry, the correlation length for the $S^z(x) S^z(0)$ sector is much smaller than that of the $S^+(x) S^-(0)$ sector.
    \item Because the state is ferromagnetic, the Fourier transform of the correlation function $C_{XY}(n) = \langle S^x(n)S^x(0) + S^y(x)S^y(0) \rangle$ does \textit{not} have a dominant peak away from $q=0$.
\end{enumerate}

In this appendix, we corroborate this claim by showing the 1D Heisenberg model exhibits this behavior. The Heisenberg Hamiltonian is
\begin{equation}
    H = J\sum_i \v{S}_i \cdot \v{S}_{i+1}.
    \label{eq:1d_Heisenburg}
\end{equation}
When $J<0$, the model is ferromagnetic and its ground state sponteneously breaks rotational symmetry. On the other hand, for $J>0$, the model is antiferromagnetic. Its ground state is a rotationally invariant $S=0$ state with gapless excitations (see e.g. \cite{fradkin2013field}). We show that the above criteria are sufficient to distinguish between these two possibilities.

\subsection{Ferromagnetic Heisenberg model}
We start in the ferromagnetic case $J=-1$. A ground state of ferromagnetic Heisenberg model is a product state $\ket{\uparrow\uparrow\cdots \uparrow\uparrow}$. Other ground states can be obtained by applying the lowering operator $S^-$ to this state. For a finite system with $N$ spins, there are $N+1$ ground states.

We now look at the ground state in infinite system via infinite DMRG. Numerically, it is standard to conserve quantum number per unit cell. In our case we can conserve $S^z$ quantum number. Taking a two-site unit cell, we have a choice of $S^z = 1$ or $S^z = 0$, corresponding to initial states $\ket{\uparrow\uparrow}^{\otimes \infty} $ or $\ket{\uparrow\downarrow}^{\otimes \infty}$, respectively. The former state is already a ground state. The latter state is not a ground state, of course, but is a good initial state for DMRG to find the ground state in the $S^z = 0$ sector.

\begin{figure}
    \centering
    \includegraphics[width=\textwidth]{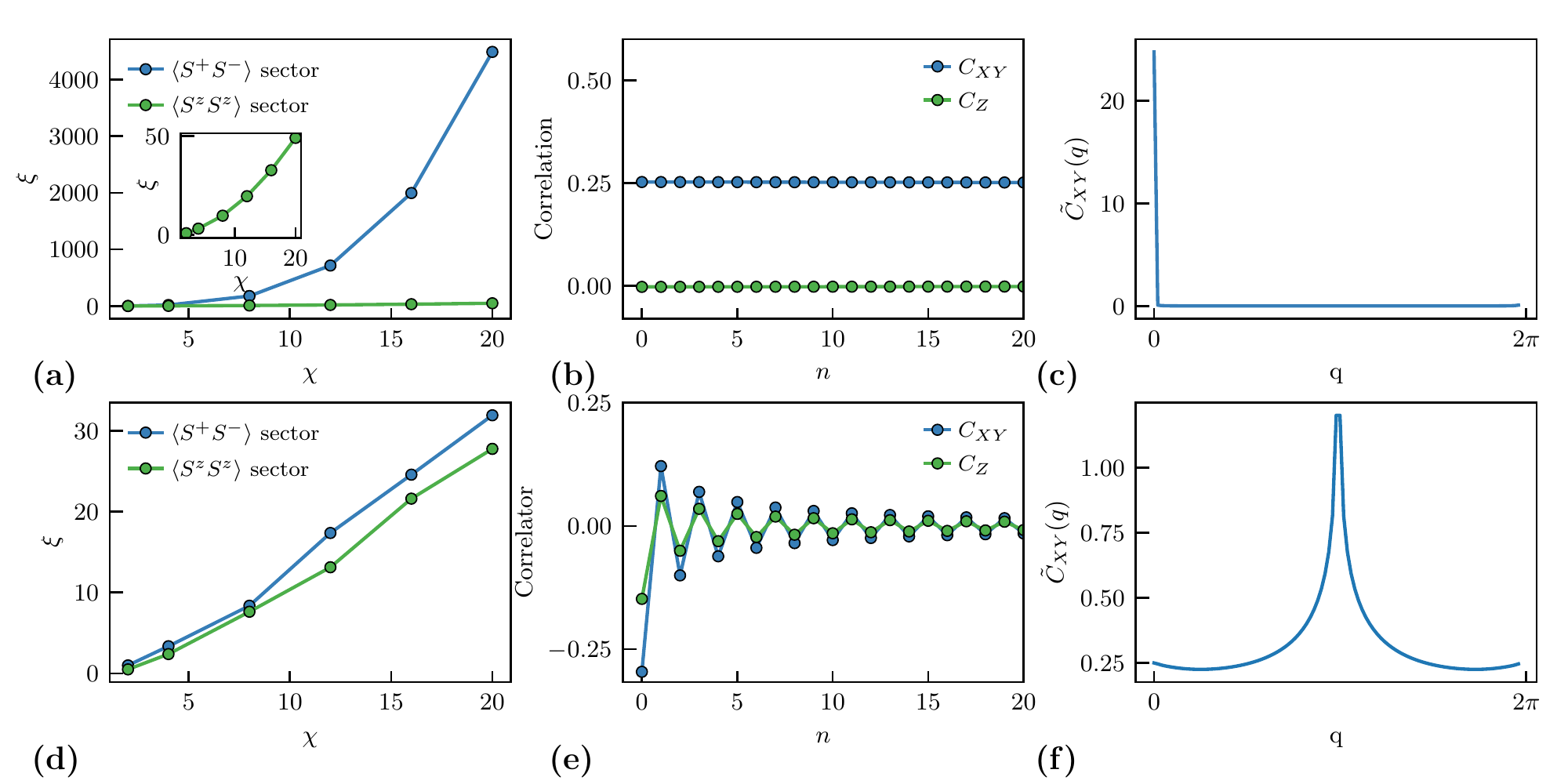}
    \caption{Correlation length and correlation functions of the 1D Heisenberg model. (a) Correlation lengths of the ferromagnetic Heisenberg model at different values of $\chi$. The inset shows a magnified view of the $S^z S^z$ correlation length. (b) $C_{XY}$ and $C_Z$ correlators of the Heisenberg model. (c) Fourier transform of the $C_{XY}$ correlator. (d)(e)(f) Same as (a)(b)(c), except for the antiferromagnetic Heisenberg model.}
    \label{fig:app_heisenberg}
\end{figure}

We now show that the $S^z = 0$ ground state of the Heisenberg model cannot be realized at any finite bond dimension. To see this, consider the following correlator:
\begin{equation}
    C_{XYZ}(n) = \langle S^z(n)S^z(0) + S^x(n)S^x(0) + S^y(n)S^y(0) \rangle = \langle S^z(n)S^z(0) + \frac{1}{2}(S^+(n)S^-(0) + S^-(n)S^+(0)) \rangle .
\end{equation}
This correlator has constant expectation value $1/4$ for the $S^z=1$ ferromagnet. Since the correlator is rotationally symmetric, we should also have $C_{XYZ} = 1/4$ in the $S^z = 0$ ground state. However, at any finite bond dimension, this correlator approaches $0$ exponentially fast as $n \to \infty$ (assuming the state is injective). This is because the connected component of correlation function asymptotically goes as $e^{-n/\xi}$ where $\xi$ is the correlation length of MPS. Since the on-site expectation values are $\langle S^z(0)\rangle = \langle S^x(0)\rangle = \langle S^y(0)\rangle$ for the $S^z = 0$ state, this means $C_{XYZ}(n)$ itself decays exponentially fast.

Therefore, in order to approximate the ground state, the correlation length $\xi$ of the MPS found from DMRG at bond dimension $\chi$ diverges rapidly as a function of $\chi$. In Fig.~\ref{fig:app_heisenberg} (a), we show the correlation length corresponding to $\langle S^+(n) S^-(0) \rangle$ and $\langle S^z(n) S^z(0)$ extracted from transfer matrix eigenvalues. The correlation length for $\langle S^+(n) S^-(0) \rangle$ diverges rapidly as a function of $\chi$, indicating in-plane ordering of spins. On the other hand, the correlation length for $\langle S^z(n) S^z(0) \rangle$ is much smaller, indicating a strong rotational symmetry breaking.

To confirm this interpretation, we measure the following in-plane and out-of-plane correlators:
\begin{align}
C_{XY}(n) &= \langle S^x(n)S^x(0) + S^y(x)S^y(0) \rangle = \frac{1}{2} \langle S^+(x)S^-(0) + S^-(x)S^+(0) \rangle, \\
C_{Z}(n) &= \langle S^z(n)S^z(0) \rangle.
\end{align}
In Fig.~\ref{fig:app_heisenberg} (b), we show the value of the correlators at $\chi=20$ as a function of $n$. The in-plane correlator $C_{XY}$ is close to constant at $1/4$, reflecting the large correlation length of $4000$, while the out-of-plane correlator $C_Z$ is close to zero. This shows that all of the spin ordering is in-plane. 

Fig.~\ref{fig:app_heisenberg} (c) shows $\tilde{C}_{XY}(q)$, the Fourier transform of $C_{XY}(n)$. There is a single dominant peak at $q=0$, confirming 1D ferromagnetic Heisenberg model satisfies the three criteria laid out above.

\subsection{Antiferromagnetic Heisenberg model}
We now look at the antiferromagnetic (AFM) Heisenberg model to contrast with the ferromagnetic scenario. As the ground state of AFM Heisenberg model has $S=0$, we can find it in the $S^z = 0$ sector with DMRG. Due to its gapless nature, there are superficial similarities in its behavior to the ferromagnetic case. In the following, we show how the criteria above can be used to distinguish it from the ferromagnetic case. 

We show in Fig.~\ref{fig:app_heisenberg}(d) the correlation length of AFM Heisenberg model as a function of $\chi$. Unlike in the ferromagnetic case, $\langle S^+(n) S^-(0) \rangle$ and $\langle S^z(n) S^z(0) \rangle$ have similar correlation length, reflecting the $SO(3)$ invariance of the ground state. A similar feature can be observed in the correlators plotted in Fig.~\ref{fig:app_heisenberg}(e); the $C_Z$ correlator is roughly the half of $C_{XY}$ correlator.

Finally, we show the Fourier transform of $C_{XY}$ correlator in Fig.~\ref{fig:app_heisenberg} (f). Reflecting the antiferromagnetic nature of spin correlation, we see the dominant peak is at $q=\pi$. These observations fully distinguish the antiferromagnetic case from the ferromagnetic case.

\section{Phase diagram analysis at the neutral sector}
\label{app:neutral_sector}

This final appendix details the phases found in the $(\tau^z,s^z) = (0,0)$ sector This is the only sector where flavor polarization is not enforced by quantum numbers, allowing a rich array of phases to appear. As a result, two challenges arise in the numerical simulation: 1) With more active degrees of freedom, the strongly correlated state cannot be represented or characterized until reaching sufficiently large bond dimensions. 2) With close competition between different low-energy states, including flavor-polarized and unpolarized ones, DMRG ground states are more likely to exhibit a mixture of several orders at finite bond dimensions.

\begin{figure}
    \centering
    \includegraphics[width=\textwidth]{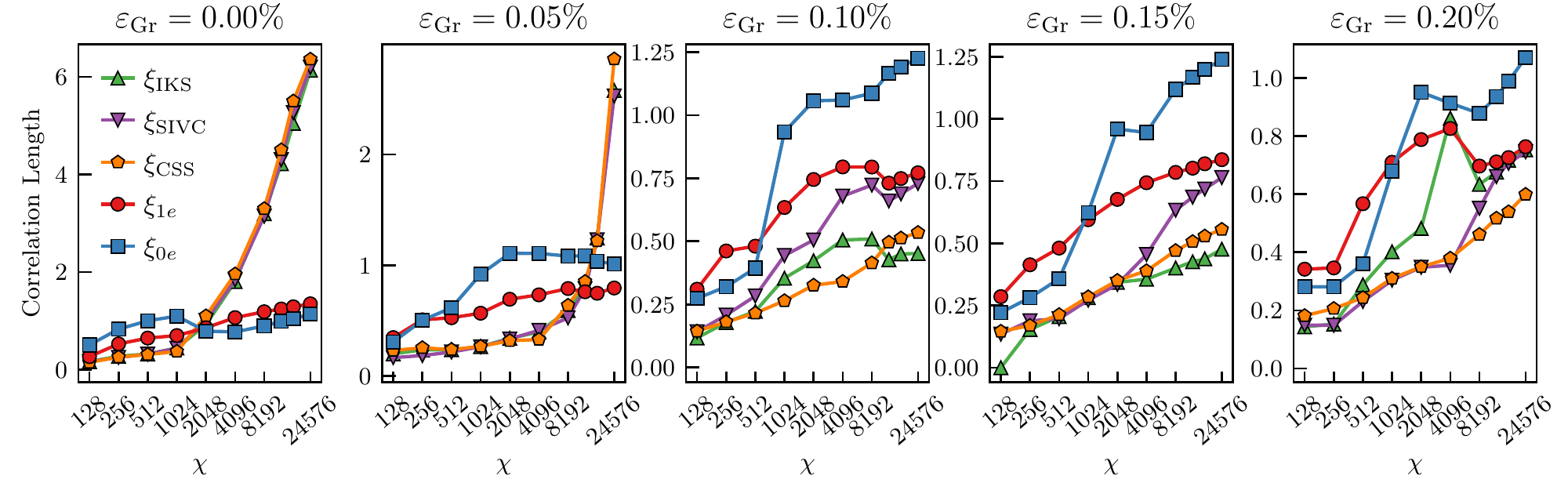}
    \caption{The correlation lengths in the $(\tau^z, s^z) = (0,0)$ sector as a function of bond dimension $\chi$. Here $\xi_{0e}$ corresponds to neutral (density-density) correlations, $\xi_{1e}$ is the maximum electron-electron correlation length among all momentum sectors, and $\xi_{\rm{IVC}}, \xi_{\rm{SIVC}}$, and $\xi_{\rm{CSS}}$ correspond to $\hat{\Delta}_{\rm{IKS}}$, $\hat{\Delta}_{\rm{SIVC}}$, and $\hat{\Delta}_{\rm{CSS}}$ correlations respectively (defined in the text).}
    \label{fig:app_correlation_lengths_00_sector}
\end{figure}

As an overview, Fig.\,\ref{fig:app_correlation_lengths_00_sector} displays the correlation lengths categorized by charge sectors for each state at different strains $\eg$ and bond dimensions $\chi$. Accordingly, we will break down our phase diagram discussion into three non-disjoint regions:

\begin{enumerate}
    \item At $\eg\ge0.1\%$, the states feature leading correlations in both neutral ($0e$) and one-electron ($1e$) charge sector, signaling a \textbf{symmetric metal}. We will take $\eg=0.2\%$ as an example to illustrate the characteristics of this order.
    \item At $\eg\le 0.05\%$, the states feature leading correlations in inter-valley and inter-spin sectors, much stronger than neutral/one-electron correlation at large $\chi$, signaling \textbf{spontaneous flavor polarization}. We will confirm that the states at $\eg=(0,0.05)\%$ are each consistent with the spin-polarized QAH-IVC and IKS order discussed in previous sections.
    \item At $\eg=(0.1,0.15)\%$, the states also feature a VDW/SDW order coexisting with the metallic order, signaling a \textbf{mixed order} (unlike the case at $\eg=0.2\%$). This indicates close competition between different orders in the regime of intermediate strain.
\end{enumerate}

We note that the states in this sector may only be the overall ground state of the system for $\eg > 0.2\%$, a fact we return to at the end of this appendix.

\subsection{Symmetric (``Normal") Metal at high $\eg$}
\label{app:symmetric_metal}

\begin{figure}
    \centering
    \includegraphics[width=\textwidth]{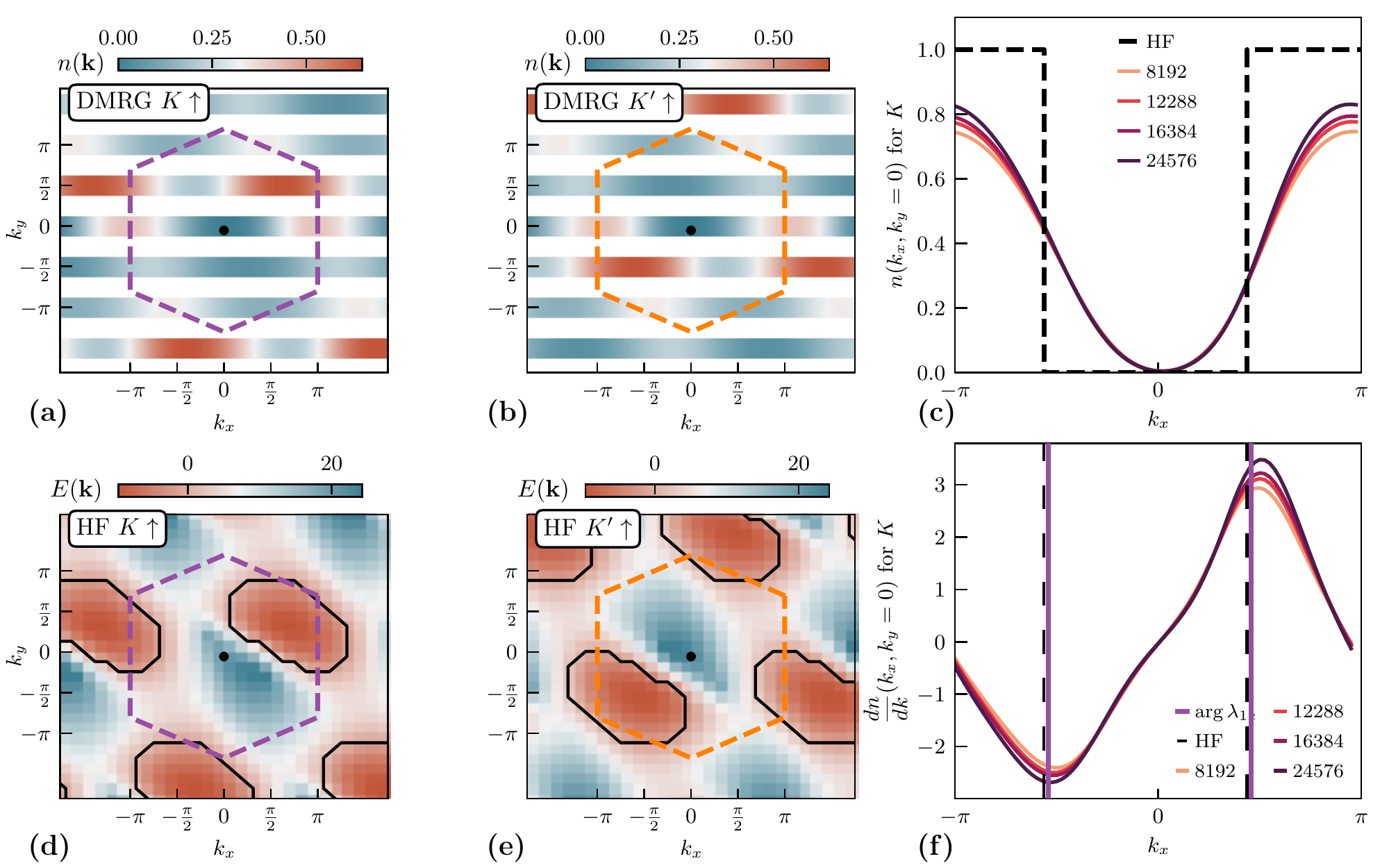}
    \caption{The symmetric metal in the $(0,0)$ sector at large strain $\eg =0.2\%$.
    (a,b) Electron occupations within DMRG in the $(K, K')$ valleys respectively. The state has spin flip $s^x$ symmetry, so the $\downarrow$ occupations are identical. Note that the region around $\Gamma$ (black dot) is fully depleted.
    (c,d) Energies of the normal metal within fully-symmetric Hartree-Fock. Energies $E(k)$ 
    (e) The DMRG electron occupations in the $K$ valley along the $k_y =0$ wire at various bond dimensions. Dashed black lines give the corresponding occupations in SCHF. One can see the DMRG occupations are slowly increasing outside of the nascent ``Fermi pocket".
    (f) Derivatives of the electron occupations $dn/dk$. One can see the occupations are increasing most quickly near the ``Fermi surfaces". Also shown are $q_x = \arg \lambda_{1e}$, the momentum associated to the $1e$ correlation channel in the system. Note that it matches the SCHF Fermi surfaces extremely well.
    Parameters: $\kappa =0.65, \eg =0.2\%, L_y = 4$, $\chi=24576$ for (a-b).}
    \label{fig:app_symmetric_metal}
\end{figure}

We start with a discussion of the symmetric or ``normal" metal state: a metallic state where all discrete and continuous symmetries are preserved, with Fermi surfaces in all four Fermion flavors. This phase is the natural ground state in the limit of weak interactions; it is what results from the non-interacting band structure to filling $\nu=-3$. Fermi liquid theory tells us that metals are relatively stable in the presence of interactions. The normal metal is therefore a reasonable phase to appear in the intermediate coupling regime. Unfortunately, we shall see that it is inherently difficult to resolve in DMRG.

The zeroth order expectation is that one starts with non-interacting dispersion in Eq.\eqref{eq:active_band_many_body_Hamiltonian} and fills the lower band to $1/4$ filling. Due to the large Hartree peak near $\Gamma$ as shown in Fig.\,\ref{fig:app_hartree_peak}(c), one expects Fermi surface(s) which do \textit{not} enclose $\Gamma$, i.e. the normal metal depletes the $\Gamma$ point just as the IKS and CSS phases do. 

To incorporate the effect of interactions, we first work at the mean-field level to understand the phenomenology. We performed self-consistent Hartree-Fock (SCHF) at $\nu=-3$ at size $16\times 16$ with all $8$ active bands. By enforcing all symmetries, we stabilize a self-consistent normal metal. Fig.\,\ref{fig:app_symmetric_metal}(d,e) shows the Hartree-Fock bandstructure of the lowest (partially filled) band in both valleys, as well as the sharp Fermi surfaces within SCHF. As expected, the region near $\Gamma$ is unoccupied in both valleys.

We now turn to DMRG. As gapless systems, metals are inherently difficult to capture within DMRG, and a 2D Fermi surface with four Fermion species is even harder. Fig.\,\ref{fig:app_symmetric_metal} shows the electron occupations in both valleys at $\chi=24576$, the largest accessible bond dimension. The electron occupations manifestly respect time-reversal and spin-flip $s^x$ symmetry. Furthermore, electrons are fully depleted near $\Gamma$ in both valleys, and Fermi (electron) pockets closely match the Fermi surfaces from SCHF. We can even identify nascent ``Fermi surfaces" forming in the electron occupations, where the derivative $dn/dk$ is peaked and increases with $\chi$ [Fig.\,\ref{fig:app_symmetric_metal}(c,f)]. The location of the Fermi surfaces agrees well with the ones predicted by SCHF, and also coincides with the corresponding eigenvalues $\lambda_{1e}$ of the one-electron sector of the transfer matrix [Fig.\,\ref{fig:app_symmetric_metal}(f)]. Therefore the occupations and general phenomenology are consistent with the expectation for a normal metallic state. 

The observed correlation lengths are also consistent with a normal metal, albeit not wholly conclusive. Recall that interacting Fermi liquids in 1d are described by $c=1$ Luttinger liquids via bosonization. Both particle-hole excitations and charge $\pm$ excitations are gapless, and the occupations near the ``Fermi surface" are power-law rather than a step function like in higher dimensions. A quasi-1D cylinder can be thought of as a coupled wire construction with $L_y$ wires, each containing $N_f$ flavors. In our case, this gives a model of $c=N_f L_y = 16$ coupled Luttinger liquids. Recall that the entanglement entropy divergences with the correlation length as  \cite{pollmann2009_finite_entanglement}
\begin{equation}
    S = \frac{c}{6} \log \xi = \frac{1}{\sqrt{\frac{12}{c}}+1} \log \chi
\end{equation} when representing gapless systems.  This large central charge means the normal metal is extremely challenging to capture within DMRG. We indeed see apparently-diverging correlation lengths in both the $0e$ and $1e$ sectors. We identify the source of the correlations $(q_x,q_y)$ in the Brillouin zone using the phase information and symmetry sector as in the IKS phase described above. We see that the $1e$ correlations come from the vicinity of the SCHF Fermi surface, and the $0e$ correlations come from the vectors \textit{between} the nascent Fermi surfaces. Using the alternative form $S \propto \log \chi$, we find that the correlations are consistent with $c=16$, but even the largest accessible bond dimension of $\chi = 24576$ is insufficient to determine the scaling behavior precisely. Meanwhile, the intervalley and interspin correlations are subleading but still substantial in the state, which can be viewed as particle-hole excitation between different sectors. Therefore the DMRG state at $\eg = 0.2\%$ is consistent with a normal metal state in a variety of non-trivial ways. We suggest the normal metal is indeed the ground state order at $\eg = 0.2\%$.

\subsection{Ferromagnetic QAH-IVC \& IKS order at low $\eg$}
\label{app:spinfulIKS}

\begin{figure}[t]
    \centering    
    \includegraphics[width=\textwidth]{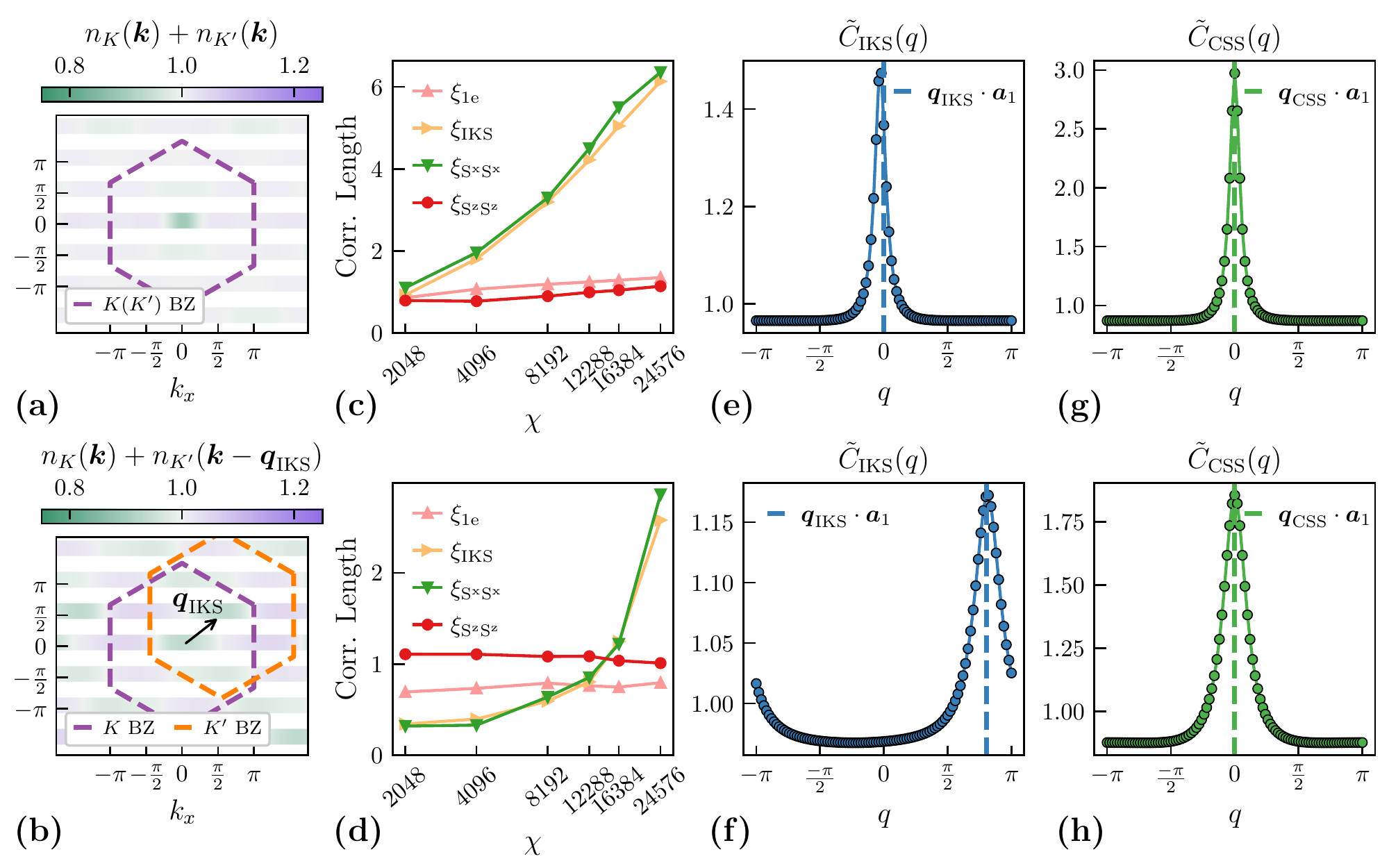}
    \caption{The spinful QAH-IVC (a,c,e,g) and IKS (b,d,f,h) phases. (a-b) The total electron density of each phase in Eq.\eqref{eq:IKS_invariant_occupations}, where (b) undergoes a relative boost by $\qIKS$. (c-d) Correlation lengths of various sectors as a function of $\chi$. The IVC correlation length, $\xi_\mathrm{IKS}$, and spin-flip correlation length, $\xi_\mathrm{S^xS^x}$, are larger than other sectors. (e-f)  The Fourier transform of $C_\mathrm{IKS}(n)$. The peak at $\qIKS\cdot\v{a}_1$ agrees with $\qIKS$ obtained from occupations $n(\v{k})$. (g-h) The Fourier transform of $C_\mathrm{CSS}(n)$ signaling spin-flip correlations, which shows a peak at $q=0$. (Unlike previous figures, $q=0$ is positioned at the center in (e-h) for a better view.)}
    \label{fig:app_spinful-QAH-IVC}
\end{figure}

This section will establish that the state at $\eg = 0\%$ has QAH-IVC order and the state at $\eg=0.05\%$ is consistent with ferromagnetic IKS order. The nature of the two states is summarized in Fig. \ref{fig:app_spinful-QAH-IVC}.

As mentioned in the main text, we identify the $\eg=0$ state as a flavor-rotation of the QAH state discussed above. The first piece of evidence is time-reversal symmetry breaking: the order parameter $C_{\rm{QAH}} = 0.87$. Together with the near-uniform occupation of the Brillouin zone, Fig \ref{fig:app_spinful-QAH-IVC}(a), we can identify this an an anomalous Hall insulator. From \ref{fig:app_spinful-QAH-IVC}(c,e) one can see the intervalley correlations are substantial, but the state has $\qIKS =0$ (i.e. no IKS ordering). Since we are in the $\tau^z=0$ sector, we identify this as the intervalley coherent version of the Hall insulator, QAH-IVC. Both this state and the $\eg=0.05\%$ state are consistent with ferromagnetic spin ordering, as we describe below.

For $\eg=0.05\%$, $C_{\rm{QAH}} = 0.003$ so time-reversal symmetry is unbroken. Fig. \ref{fig:app_spinful-QAH-IVC}(d) show that intervalley correlations are increasing quickly with $\chi$. As before, both the shifted BZ occupations in Fig. \ref{fig:app_spinful-QAH-IVC}(a) and the Fourier transform of the IKS correlation function Fig. \ref{fig:app_spinful-QAH-IVC}(d) give $\qIKS \approx (0.277,0.25)$. We therefore identify this state as having IKS order.

Finally, we use spin-spin correlations to establish the ferromagnetic ordering of the spins. As a starting note, the simplest ferromagnetic order --- spins aligned in the $+z$ direction --- is incompatible with the $(\tau^z,s^z) = (0,0)$ sector. Given that we only conserve $U(1) \subset SU(2)$, we must deduce ferromagnetism in a slightly indirect way. As discussed in App. \ref{app:heisenburg_analogy}, the hallmarks of ferromagnetic states in the $s^z=0$ sector are:
\begin{enumerate}
    \item The correlation length for $\braket{s^+(n)s^-(0)}$ diverges with $\chi$.
    \item The correlation length $\xi_{S^zS^z}$for $\braket{s^z(n)s^z(0)}$ is much smaller than $\xi_{S^xS^x}$ for $\braket{s^+(n)s^-(0)}$.
    \item The correlation function doesn't have peaks at nonzero momenta.
\end{enumerate}
The first condition establishes spin ordering, the second shows rotational symmetry breaking, and the final condition rules out antiferromagnetic order. In Figs. \ref{fig:app_spinful-QAH-IVC}(c,d), we see that the correlation length $\xi_{S^xS^x}$ are increasing rapidly with $\chi$, establishing the first property. (This correlation length corresponds to $C_{CSS}$, which is defined below Eq. \eqref{eq:CSS_operator}.) We see there is no corresponding increase in $\xi_{S^zS^z} = \xi_{0e}$, establishing property 2. (We note that the corresponding correlation function ``$\braket{s^z(n)s^z(0)}$" cannot be directly computed due to spin conservation in both valleys.) Finally the Fourier transform of $C_{CSS}$ shows a single peak at $q=0$ in Fig. \ref{fig:app_spinful-QAH-IVC}(g,h), establishing condition 3. Combining these observations, we identify both $\eg=0,0.05\%$ as having ferromagnetic order alongside their QAH-IVC and IKS orders in the sublattice/valley flavors.

\subsection{``Mixed" order at intermediate $\eg$}
\label{app:mixed_order}

The $\eg = (0.1,0.15)\%$ states of the $(\tau^z,s^z) = (0,0)$ sector feature what we will term ``mixed" order, with features of both normal metal and flavor-polarized states, complicated by possibly-transient translation-breaking order. We start with Fig.\,\ref{fig:app_correlation_lengths_00_sector} showing the correlation lengths at these strain values. One can see their leading correlations in ${0e}$ and ${1e}$ sectors, consistent with a normal metal. However, they also manifest strong spin asymmetry in intervalley correlations, namely with $\xi_{\rm{SIVC}}$ (between opposite spins) much larger than $\xi_{\rm{IVC}}$ (between the same spins) in both states. This behavior deviates from the observation in either the normal metal or the ferromagnetic orders, which we hypothesize to result from ``partial" spin polarization to be further examined.

Next, we take a closer look at the pattern of flavor polarization in these states, which reveals a strong spin density wave (SDW) order and a transient valley density wave (VDW) order. These orders break translation $T_{\v{a}_1}$ along the cylinder down to $2T_{\v{a}_1}$, and are detected by order parameters that probe flavor-resolved charge imbalances between adjacent rings on the cylinder: Let $-1/2 \le P(k_y)_{\sigma,\tau} < 1/2$ be the polarization (center of charge) of the hybrid Wannier orbitals \cite{Soejima2020_efficient}, then the order parameters are defined as
\begin{align}
    N_\mathrm{VDW}(\pi) &= \frac{1}{L_x}\sum_{n,k_y,\sigma\tau s} e^{i\pi n P(k_y,\sigma,\tau)} \langle c^\dagger_{n,k_y,\sigma\tau s} [\tau^z]_{\tau \tau} c_{n,k_y,\sigma\tau s} \rangle, \\
    N_\mathrm{SDW}(\pi) &= \frac{1}{L_x}\sum_{n,k_y,\sigma\tau s} e^{i \pi n P(k_y,\sigma,\tau)} \langle c^\dagger_{n,k_y,\sigma\tau s} [s^z]_{s s} c_{n,k_y,\sigma\tau s} \rangle.
    \label{eq:VDW_SDW_correlators}
\end{align}
The sum is taken over some segment with unit cells $n=1,2,\dots,L_x$, and here we choose $L_x$ even since the density waves double the unit cell.

\begin{figure}
    \centering
    \includegraphics[width=\textwidth]{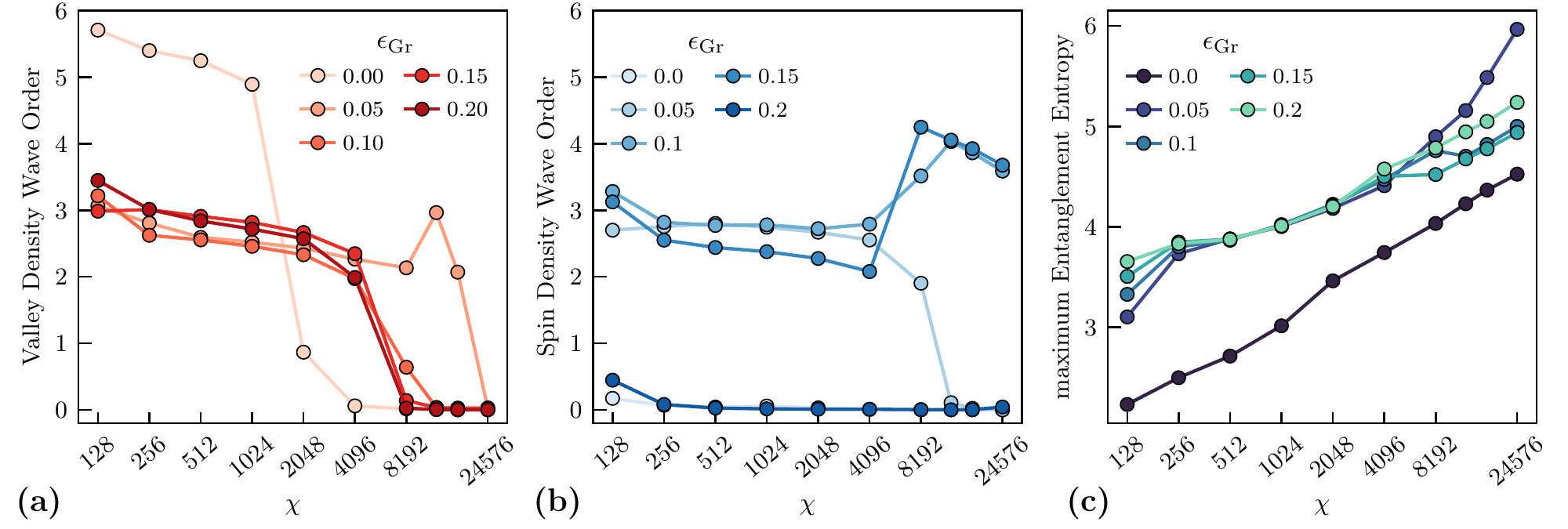}
    \caption{Other characteristics of $(\tau^z, s^z) = (0,0)$ states at different $\eg$ as a function of bond dimension $\chi$. (a) Valley density wave measured by $N_\mathrm{VDW}$; (b) Spin density wave measured by $N_\mathrm{SDW}$; (c) Entanglement entropy.}
    \label{fig:app_mixed_order}
\end{figure}

Fig.\,\ref{fig:app_mixed_order}(a-b) show the behavior of VDW and SDW orders in the ground states at all strain values. In all cases, the VDW order is present at smaller bond dimensions $\chi$ but disappears at the largest $\chi=24576$. Note that they share a common trend of ``finite-$\chi$ phase transition": $N_{\rm{VDW}}$ has a substantial value, perhaps decreasingly slightly, until it abruptly vanishes. 
As for the SDW order, at $\eg = (0, 0.05, 0.2)\%$ it has vanished by $\chi=16384$, but at $\eg=(0.1,0.15)\%$ it persists even at the largest $\chi$. This is consistent the hypothesis of ``partial polarization" from correlation lengths: the SDW order results in nonzero $s^z$ polarization alternating in each unit cell (thus not violating the net $s^z=0$), arguably leading to the spin asymmetry found in intervalley correlations. Meanwhile, unlike ferromagnetic orders, there is no net spin polarization either in the $z$ direction or in the $xy$ plane, which agrees with the weak spin-spin correlations. Nevertheless, we caution that the SDW order could also be a finite-$\chi$ effect, which may disappear suddenly if one access even larger bond dimensions.

For completeness, we also comment on some other characteristics of the ``mixed" order. Fig.\,\ref{fig:app_mixed_order}(c) includes the entanglement entropy of the ground states at different $\eg$, where the states with ``mixed" order show similar growth in entanglement as other phases. We also find that no choice of $\qIKS$ makes their electron occupations entirely uniform across the Brillouin zone, though certain choices that offset the Hartree peaks work somewhat well.

Given the mix of signals for $\eg = (0.1,0.15)\%$ states --- metallic order, spin-asymmetric correlations, and possibly transient spin density waves --- we do not make a definitive claim of the phase at intermediate $\eg$, and instead refer to these states as ``mixed" order on phenomenological grounds. We note that these states has very close energy to the spin-polarized IKS states in the (0,1) sector (as shown in Fig.\,\ref{fig:phase_diagram}(a)), suggesting the close competition between different orders in this regime.

\subsection{Higher Strain and the Phase Diagram at $\nu=-3$}
\label{app:overall_phase_diagram}

\begin{figure}
    \centering
    \includegraphics[width=0.45\textwidth]{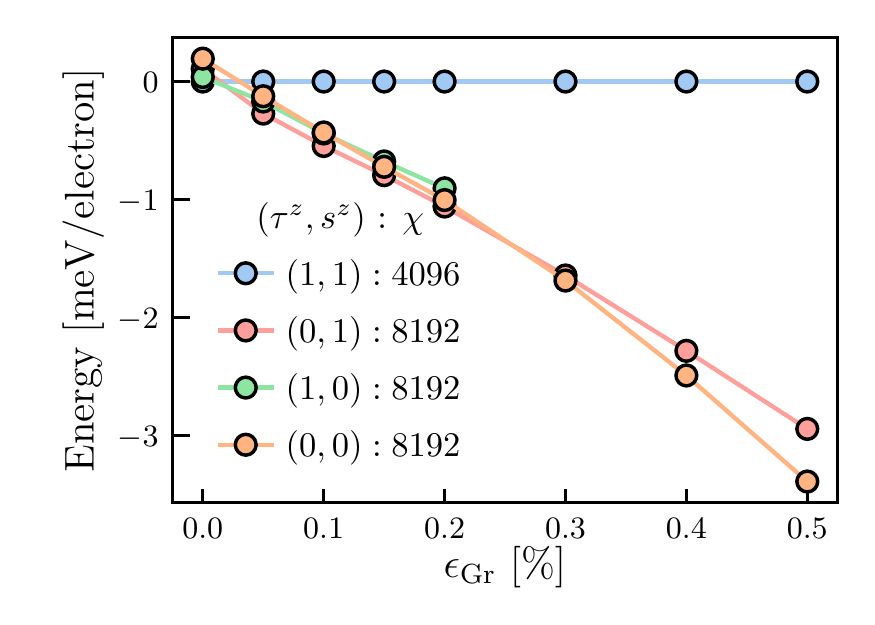}
    \caption{DMRG energies in all four sectors as a function of heterostrain $\eg$. Due to computational expense, we use $\chi=8192$ except in the $(1,1)$ which is already well-converged by $\chi=4096$. The $(1,0)$ sector was not computed beyond $\eg=0.2\%$, but is expected to be above $(0,0)$ and $(0,1)$ in energy. Note that the $(0,0)$ is clearly the lowest energy for $\eg > 0.2\%$. The phases and transitions are discussed in the text in detail.}
    \label{fig:app_energy_comparison_large_strain}
\end{figure}

We conclude with some data on higher strain values and speculation on the true phase diagram. Fig.\,\ref{fig:app_energy_comparison_large_strain} gives DMRG ground state energies in all four sectors up to $\eg = 0.5\%$. Due to numerical expense, we restrict to $\chi=8192$ or less. Recall in Fig.\,\ref{fig:phase_diagram}(a) of the main text that $(0,0)$ [normal metal] is the global ground state at $\eg = 0.2\%$, with an energy difference of $<\,\SI{0.01}{\milli\electronvolt}$ from the $(0,1)$ [IKS] state. For even larger strains, we see the $(0,0)$ sector is the clear energetic winner. As more strain continues to increase the bandwidth of the flat bands, this is physically reasonable. We note, however, that our assumptions that the active bands are relatively well-isolated from the remote bands likely breaks down for sufficiently large strain.

Taking the identifications of states in our model at face value and incorporating all sectors, we arrive at the following possible phase diagram of $\nu=-3$.
\begin{enumerate}
    \item Quantized anomalous hall order at $\eg \le 0.05\%$.
    \item IKS order at $0.05\% < \eg < 0.2\%$.
    \item A normal metal at $0.2\% \le \eg$.
\end{enumerate}
This is shown in Fig.\,\ref{fig:overview}(d). We note that the exact phase boundaries are expected to change somewhat depending on finite-size effects and other parameters such as $\kappa$. For instance, using $L_y =6$ stabilizes QAH at a larger strain value (as does decreasing $\kappa$ to approach the chiral limit). Furthermore, the energy differences between these phases are only slightly larger than the numerical precision of our Hamiltonian and ground state --- and much larger than the physical uncertainty in our model. Nevertheless, we expect the basic picture of the three phases to be robust.

Finally, we comment on the phase transitions. As QAH and IKS order lie in different quantum number sectors, we expect a first-order phase transition between them. For the $(0,0)$ sector, the ferromagnetic IKS order breaks symmetries, whereas the normal metal does not. The transition between them may in principle be second-order (though we do not rule out a first-order transition). If this is the case, any quantum critical behavior would manifest in the region where we have found the ``mixed" order.

\end{document}